\documentclass[final]{siamltex}
\usepackage{epsfig}
\usepackage{amsmath}
\usepackage{amssymb}
\usepackage{setspace}
\usepackage{epsfig}
\usepackage{graphicx}
\usepackage{amsmath}
\usepackage{amssymb}
\usepackage{setspace}

\usepackage{algpseudocode}
\usepackage[ruled,linesnumbered,vlined]{algorithm2e}

\renewcommand{\algorithmicrequire}{\textbf{Input:}}  
\renewcommand{\algorithmicensure}{\textbf{Output:}} 

\newtheorem{defn}{\noindent $\mathbf{Definition}$}[section]

\newtheorem{thm}[defn]{$\mathbf{Theorem}$}

\title{QCMC: Quasi-conformal Parameterizations for Multiply-connected domains}
\author{Kin Tat Ho and Lok Ming Lui}

\begin{document}

\maketitle

\begin{abstract}
This paper presents a method to compute the {\it quasi-conformal parameterization} (QCMC) for a multiply-connected 2D domain or surface. QCMC computes a quasi-conformal map from a multiply-connected domain $S$ onto a punctured disk $D_S$ associated with a given Beltrami differential. The Beltrami differential, which measures the conformality distortion, is a complex-valued function $\mu:S\to\mathbb{C}$ with supremum norm strictly less than 1. Every Beltrami differential gives a conformal structure of $S$. Hence, the conformal module of $D_S$, which are the radii and centers of the inner circles, can be fully determined by $\mu$, up to a M\"obius transformation. In this paper, we propose an iterative algorithm to simultaneously search for the conformal module and the optimal quasi-conformal parameterization. The key idea is to minimize the Beltrami energy subject to the boundary constraints. The optimal solution is our desired quasi-conformal parameterization onto a punctured disk. The parameterization of the multiply-connected domain simplifies numerical computations and has important applications in various fields, such as in computer graphics and vision. Experiments have been carried out on synthetic data together with real multiply-connected Riemann surfaces. Results show that our proposed method can efficiently compute quasi-conformal parameterizations of multiply-connected domains and outperforms other state-of-the-art algorithms. Applications of the proposed parameterization technique have also been explored.
\end{abstract}
\begin{keywords}
Quasi-conformal, parameterization, multiply-connected, Beltrami differential, conformal module, Beltrami energy
\end{keywords}

\pagestyle{myheadings}
\thispagestyle{plain}
\markboth{Ho and Lui}{Quasi-conformal parameterizations for multiply-connected domains}

\section{Introduction}\label{introduction}
Parameterization refers to the process of mapping a complicated domain one-to-one and onto a simple canonical domain. For example, according to the Riemann mapping theorem, a simply-connected open surface can be conformally mapped onto the unit disk $\mathbb{D}$. The geometry of the canonical domain is usually much simpler than its original domain. Hence, by parameterizing a complicated domain onto its simple parameter domain, a lot of numerical computations can be simplified.

Parameterizations have been extensively studied and various parameterization algorithms have been developed. In particular, conformal parameterizations have been widely used, since it preserves the local geometry well. For example, in computer graphics, conformal parameterizations of 3D surfaces onto 2D images have been applied for texture mapping \cite{Levy:2002}. While in medical imaging, conformal parameterizations have been used for obtaining surface registration between various anatomical structures, such as the brain cortical surfaces \cite{Gu1,Gu2,LuiBHFHP}. Conformal parameterizations have also been applied to solve PDEs on complicated 2D domains or surfaces \cite{LuiPDE}\cite{Krichever}\cite{Krichever2}.
 
In case of extra constraints have to be enforced, obtaining conformal surface parameterizations may not be feasible. In such situation, quasi-conformal parameterizations, which allow bounded amount of conformality distortions, have to be considered. The conformality distortion can be measured by {\it Beltrami differential}. Quasi-conformal parameterization of a complicated domain onto a simple parameter domain is useful and have found important applications in various fields. For example, in computer graphics, constrained texture mapping that matches feature landmarks are quasi-conformal parameterizations \cite{LuiTMap}. Besides, by parameterizing two surfaces quasi-conformally onto simple parameter domains in $\mathbb{R}^2$, quasi-conformal map between two Riemann surfaces with a given Beltrami differential can be easily computed. Quasi-conformal parameterization can also simplify the process of solving the elliptic partial differential equations (PDEs) on a complicated surface. Elliptic PDEs arise in many imaging problems, such as in surface registration. Through quasi-conformal parameterization, the elliptic PDE on a complicated domain can be formulated into a simpler PDE on the simple parameter domain. For instance, the elliptic PDE $\nabla \cdot (A\nabla f) = g$ on a multiply-connected surface $S$ with certain boundary conditions can be converted to a simpler PDE: $\Delta f\circ \phi = g\circ \phi$ on a circle domain, where $\phi$ is the quasi-conformal parameterization of $S$ whose Beltrami differential is given by $A$ (SPD matrix). The simpler PDE defined on a simpler parameter domain can be solved much easier. Other applications of quasi-conformal parameterizations include remeshing, grid generation, texture mapping, spline fitting and so on. Because of its wide applications, various algorithms for quasi-conformal parameterizations have been proposed recently \cite{Mastin2,Daripa,LuiBHF,LuiQuasiYamabe}.

Most parameterization algorithms deal with domains or surfaces with simple topologies, such as simply-connected open surfaces. Parameterizing domains with complicated topologies is generally challenging. In this work, our main focus is to compute the quasi-conformal parameterization (QCMC) of the multiply-connected domain onto the punctured disk (a unit disk with several inner disks removed). According to Quasi-conformal Teichm\"uller theories, every quasi-conformal map is associated with a Beltrami differential, which is a complex-valued function defined on the source domain with supremum norm strictly less than 1. The Beltrami differential measures the conformality distortion. Given a Beltrami differential $\mu$, a multiply-connected domain can be parameterized quasi-conformally onto a punctured disk. The inner radii and centers of the punctured disk depend on the Beltrami differential. We propose an iterative algorithm to simultaneously look for the conformal module and the quasi-conformal parameterization of a multiply-connected 2D domain or surface. The key idea is to minimize the Beltrami energy subject to boundary constraints. By incorporating the conformal module into the energy functional of the optimization problem, the quasi-conformal map together with the conformal module can be simultaneously optimized. In particular, when $\mu$ is set to be zero, a least square conformal map (LSCM) from a multiply-connected domain to a punctured disk can be obtained. Experiments have been carried out on synthetic data together with real multiply-connected Riemann surfaces. Results show that our proposed method can efficiently compute quasi-conformal map associated to a given Beltrami differential and outperforms other state-of-the-art algorithms. Applications of the proposed parameterization technique have also been explored.

The rest of the paper is organized as follows. In Section \ref{related}, we describe some previous works closely related to this paper. In Section \ref{background}, we describe some basic mathematical concepts. Our proposed model is explained in details in Section \ref{proposed}. The numerical implementation details will be described in Section \ref{implementation}. In Section \ref{experiment}, we show some experimental results of the proposed method. Applications of the proposed parameterization technique will be explored in Section \ref{application}. The paper is concluded in Section \ref{conclusion}.

\section{Related works}\label{related}
Parameterization has been widely studied and different parameterization algorihtms have been developed. The goal is to map a 2D complicated domain or 3D surface onto a simple parameter domain, such as the unit sphere or 2D rectangle. In general, 3D surfaces are not isometric to the simple parameter domains. As a result, parameterization usually causes distortion. Tannenbaum et al. \cite{IsoMap} proposed to obtain a close-to-isomtric parameterization, called the IsoMap, which minimizes geodesic distance distortion between pairs of vertices on the mesh. 
Eck et al. \cite{Eck} propose the discrete harmonic map for mesh parameterization, which approximates the continuous harmonic map by minimizing a metric dispersion criterion. Dominitz et al. \cite{Dominitz} proposed a parameterization, which is as area-preserving as possible, for texture mapping via optimal mass transportation. Graph embedding of a surface mesh has also been studied by Tutte \cite{Tutte}. The parameterization technique, which is now called the Tutte's embedding, was introduced. The bijectivity of the parameterization is mathematically guaranteed. Floater \cite{Floater} improved the quality of the parameterization by introducing specific weights, in terms of area deformations and conformality.

Besides, conformal parameterization has been extensively studied  \cite{Haker,Fischl2,Gu1,Gu2,Gu3,Hurdal,conformal1,conformal2,conformal3,spectralconformal}. Levy et al. \cite{Levy:2002} proposed to compute the least square confomal parameterization through an optimiziation approach, which is based on the least square approximation of the Cauchy-Riemann equations. Hurdal et al. \cite{Hurdal} proposed to compute the conformal parameterizations using circle packing and applied it to register human brains. Porter \cite{conformal1} proposed to compute the conformal maps of simply-connected planar domains by the interpolating polynomial method. Gu et al. \cite{Gu1,Gu2,Gu3} proposed to compute the conformal parameterizations of Riemann surfaces for the purpose of registration using harmonic energy minimization and holomorphic 1-forms. Later, the curvature flow method to compute conformal parameterizations of high-genus surfaces onto their universal covering spaces was also proposed, which deforms the Riemmannian metric to the uniformization metric \cite{JinRicci,InverseDistance,LuiVSRicci}. Curvature flow method can also be used to parameterize multiply-connected domains onto the punctured disks \cite{LuiCM,WangRicciMC}. Hale et al. \cite{conformal2} proposed to compute conformal maps to multiply-slit domains by using a Schwarz-Christoffel formulation. DeLillo et al. \cite{conformal3} proposed a numerical method to compute the Schwarz-Christoffel transformation for multiply-connected domains. Crowdy \cite{CrowdySC} proposed a formula for the generalized Schwarz-Christofeel conformal mapping from a bounded multiply-connected circular domain to an unbounded multiply-connected polygonal domain. Conformal maps have been widely used since it preserves the local geometry well.

Sometimes, when further constraints are enforced, exact conformal parameterizations may not be achievable. In this case, quasi-conformal parameterizations have to be considered. Recently, various algorithms for quasi-conformal parameterizations have been developed. For example, Mastin et al. \cite{Mastin2} proposed a finite difference scheme for constructing quasi-conformal mappings for arbitrary simply and doubly-connected region of the plane onto a rectangle. In \cite{Daripa}, Daripa proposed a numerical construction of quasi-conformal mappings in the plane by solving the Beltrami equation. This method was further extended to compute the quasi-conformal map of an arbitrary doubly connected domain with smooth boundaries onto an annulus \cite{Daripa2}. All of these methods deal with simple domains in the complex plane. Recently, surface quasi-conformal maps have also been studied. Lui et al. \cite{LuiBHFHP} proposed to compute quasi-conformal registration between hippocampal surfaces which matches geometric quantities (such as curvatures) as much as possible. A method called the Beltrami holomorphic flow was used to obtain the optimal Beltrami coefficient associated to the registration \cite{LuiBHF,LuiCompression,LuiBeltramirepresentation}. Wei et al. \cite{Weiface} also proposed to compute quasi-conformal mapping for feature matching face registration. The Beltrami coefficient associated to a landmark points matching parameterization was approximated. However, either exact landmark matching or the bijectivity of the mapping cannot be guaranteed, especially when very large deformations occur. In order to compute quasi-conformal mapping from the Beltrami coefficients effectively, Quasi-Yamabe method was introduced, which applied the curvature flow method to compute the quasi-conformal mapping \cite{LuiQuasiYamabe}. The algorithm can deal with surfaces of general topologies. Later, extremal quasi-conformal mappings, which minimize conformality distortion has been proposed. Lui et al. \cite{LuiTMap} proposed to compute the unique Teichm\"uller extremal map between simply-connected Riemann surfaces of finite type. The proposed algorithm was applied for landmark-based surface parameterization.

\section{Mathematical background}\label{background}
In this section, we describe briefly some basic mathematical concepts closely related to this work. For details, we refer the reader to \cite{Gardiner}.

A surface $S$ with a conformal structure is called a \emph{Riemann surface}. Given two Riemann surfaces $M$ and $N$, a map $f:M\to N$ is \emph{conformal} if it preserves the surface metric up to a multiplicative factor called the {\it conformal factor}. An immediate consequence is that every conformal map preserves angles. With the angle-preserving property, a conformal map effectively preserves the local geometry of the surface structure. 
A generalization of conformal maps is the \emph{quasi-conformal} maps, which are orientation preserving homeomorphisms between Riemann surfaces with bounded conformality distortion, in the sense that their first order approximations take small circles to small ellipses of bounded eccentricity \cite{Gardiner}. Mathematically, $f \colon \mathbb{C} \to \mathbb{C}$ is quasi-conformal provided that it satisfies the Beltrami equation:
\begin{equation}\label{beltramieqt}
\frac{\partial f}{\partial \overline{z}} = \mu(z) \frac{\partial f}{\partial z}.
\end{equation}
\noindent for some complex-valued function $\mu$ satisfying $||\mu||_{\infty}< 1$. $\mu$ is called the \emph{Beltrami coefficient}, which is a measure of non-conformality. It measures how far the map at each point is deviated from a conformal map. In particular, the map $f$ is conformal around a small neighborhood of $p$ when $\mu(p) = 0$. Infinitesimally, around a point $p$, $f$ may be expressed with respect to its local parameter as follows:
\begin{equation}
\begin{split}
f(z) & = f(p) + f_{z}(p)z + f_{\overline{z}}(p)\overline{z} \\
& = f(p) + f_{z}(p)(z + \mu(p)\overline{z}).
\end{split}
\end{equation}

\begin{figure}[t]
\centering
\includegraphics[width=3.75in]{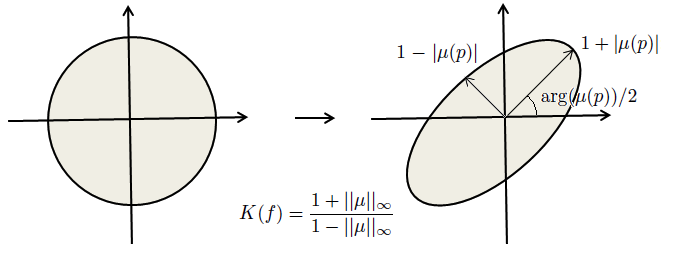}
\caption{The figure illustrates how the conformality distortion can be measured by the Beltrami coefficient.\label{fig:conformalitydistortion}}
\end{figure}

Obviously, $f$ is not conformal if and only if $\mu(p)\neq 0$. Inside the local parameter domain, $f$ may be considered as a map composed of a translation to $f(p)$ together with a stretch map $S(z)=z + \mu(p)\overline{z}$, which is postcomposed by a multiplication of $f_z(p),$ which is conformal. All the conformal distortion of $S(z)$ is caused by $\mu(p)$. $S(z)$ is the map that causes $f$ to map a small circle to a small ellipse. From $\mu(p)$, we can determine the angles of the directions of maximal magnification and shrinking and the amount of them as well. Specifically, the angle of maximal magnification is $\arg(\mu(p))/2$ with magnifying factor $1+|\mu(p)|$; The angle of maximal shrinking is the orthogonal angle $(\arg(\mu(p)) -\pi)/2$ with shrinking factor $1-|\mu(p)|$. Thus, the Beltrami coefficient $\mu$ gives us lots of information about the properties of the map (See Figure \ref{fig:conformalitydistortion}).

The maximal dilation of $f$ is given by:
\begin{equation}
K(f) = \frac{1+||\mu||_{\infty}}{1-||\mu||_{\infty}}.
\end{equation}

Suppose $f:\Omega_1\to \Omega_2$ and $g:\Omega_2\to\Omega_3$ are quasi-conformal maps, whose Beltrami coefficients are $\mu_f$ and $\mu_g$ respectively. Then, the Beltrami coefficient of the composition map $g\circ f :\Omega_1\to \Omega_3$ is given by:
\begin{equation}\label{composition}
\mu_{g\circ f} = \frac{\mu_f + r_f (\mu_g\circ f)}{1+r_f \overline{\mu_f}(\mu_g\circ f)}
\end{equation}
\noindent where $r_f = \overline{f_z}/ f_z$.

Given a Beltrami coefficient $\mu:\mathbb{C}\to \mathbb{C}$ with $\|\mu\|_\infty < 1$. There is always a quasiconformal mapping from $\mathbb{C}$ onto itself which satisfies the Beltrami equation in the distribution sense \cite{Gardiner}.

\begin{figure}[t]
\centering
\includegraphics[width=3.75in]{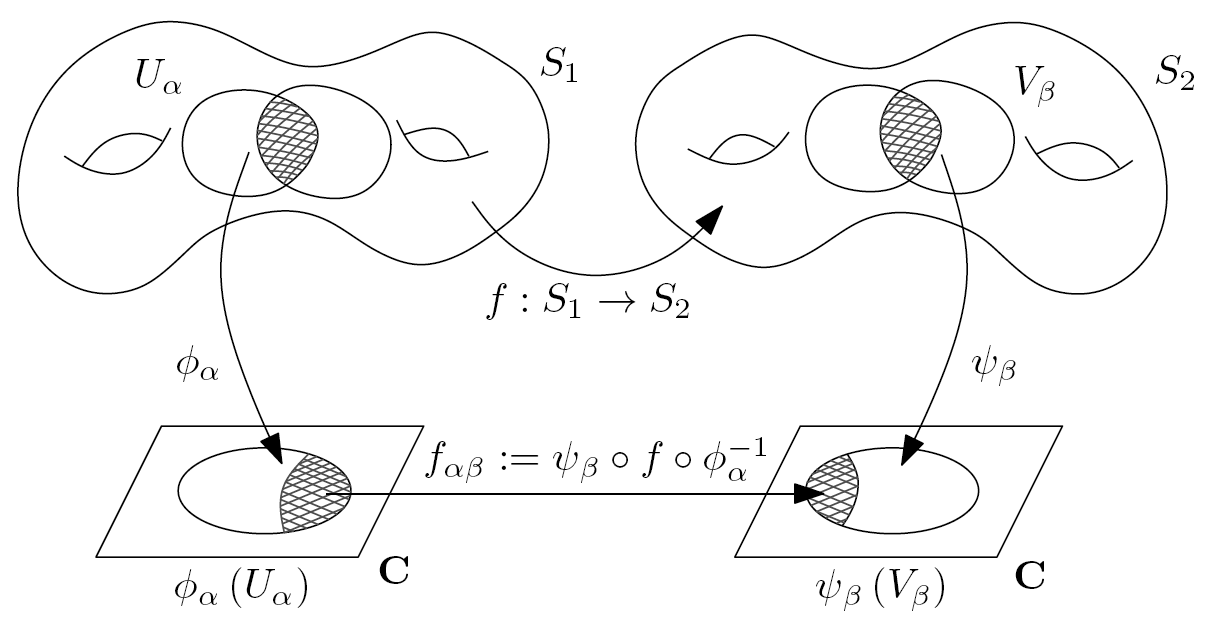}
\caption{The figure illustrates how surface quasi-conformal map is defined.\label{fig:Beltramidifferential}}
\end{figure}

Quasiconformal mapping between two Riemann surfaces $S_1$ and $S_2$ can also be defined. Instead of the Beltrami coefficient, the {\it Beltrami differential} is used. A Beltrami differential $\mu(z) \frac{\overline{dz}}{dz}$ on a Riemann surface $S$ is an assignment to each chart $(U_{\alpha},\phi_{\alpha})$ of an $L_{\infty}$ complex-valued function $\mu_{\alpha}$, defined on local parameter $z_{\alpha}$ such that
\begin{equation}
\mu_{\alpha}(z_{\alpha})\frac{d\overline{z_{\alpha}}}{dz_{\alpha}} = \mu_{\beta}(z_{\beta})\frac{d\overline{z_{\beta}}}{dz_{\beta}},
\end{equation}
\noindent on the domain which is also covered by another chart $(U_{\beta},\phi_{\beta})$. Here, $\frac{dz_{\beta}}{dz_{\alpha}}= \frac{d}{dz_{\alpha}}\phi_{\alpha \beta}$ and $\phi_{\alpha \beta} = \phi_{\beta}\circ \phi_{\alpha}^{-1}$ (See Figure \ref{fig:Beltramidifferential}).

Given a Beltrami differential $\mu(z) \frac{\overline{dz}}{dz}$ on a Riemann surface $S$, the surface can always be parameterized quasi-conformally onto a punctured disk with the prescribed Beltrami differential. The geometry (centers and radii of the inner circles) depends solely on $\mu(z) \frac{\overline{dz}}{dz}$. More precisely,

\medskip

\begin{thm}
Suppose $S$ is an open Riemann surface with multiple boundaries. Given a Beltrami differential $\mu(z) \frac{\overline{dz}}{dz}$, there exist a quasi-conformal parameterization $f:S \rightarrow D_S$ with the prescribed Beltrami differential, where $D_S$ is a unit disk with circular holes (or the punctured disk). Two such kind of parameterizations differ by a M\"obius transform .
\end{thm}

\medskip

In this work, our goal is to develop an effective numerical algorithm to compute such quasi-conformal parameterization.

\section{Proposed method}\label{proposed}
The problem we address in this paper is to compute the quasi-conformal parameterization of a multiply-connected domain $S$ with a given Beltrami differential $\mu$.

Without loss of generality, we may assume $S$ to be an open connected domain in $\mathbb{R}^2$. Thus, we can regard the Beltrami differential $\mu$ as the Beltrami coefficient. More specifically, if $S$ is a multiply-connected surface in $\mathbb{R}^3$, we can easily map it onto an arbitrary 2D domain $\Omega$ using, for instance, the IsoMap \cite{IsoMap}, the spectral conformal map algorithm \cite{spectralconformal} or the Beltrami holomorphic flow algorithm with free boundary condition \cite{LuiBHF,LuiExtremalMC}. Of course, the parameter domain $\Omega$ has an arbitrary shape (see Figure \ref{fig:conformalembed}). Denote the initial parameterization by $\phi: S\to \Omega$, whose Beltrami differerential is $\mu_{\phi}$. Our goal is to find a quasi-conformal parameterization $f:\Omega\to D_S$ between the 2D domain $\Omega$ and the punctured disk $D_S$, such that the composition map has Beltrami differential equals to $\mu$. According to equation (\ref{composition}), $f$ should be a quasi-conformal map with Beltrami coefficient $\nu$ equals to:
\begin{equation}\label{nu}
\nu = \frac{1}{r_f}\left[\frac{\mu-\mu_f}{1-\mu \overline{\mu_{\phi}}}\right].
\end{equation}

With this setting, we can now mathematically formulate our problem as follows: given a multiply-connected domain $S$ and a Beltrami coefficient $\mu:S\to \mathbb{C}$, we look for a punctured disk :
\begin{equation}\label{originalproblem1}
D_S = \mathbb{D}\setminus \bigcup_{i=1}^n B_{r_i}(c_i).
\end{equation}
\noindent where $B_{r_i}(c_i) = \{x\in \mathbb{D}: |x-c_i|< r_i\}$, $\bigcup_{i=1}^n B_{r_i}(c_i) \subset \mathbb{D}$ and $\bigcap_{i=1}^m B_{r_i}(c_i) = \phi$ ($0<r_i<1$ and $c_i\in \mathbb{D}\subset \mathbb{C}$), together with a quasi-conformal parameterization $f:S\to D_S$ such that:
\begin{equation}\label{originalproblem2}
\frac{\partial f}{\partial \bar{z}} = \mu \frac{\partial f}{\partial z}.
\end{equation}

Let $\vec{r} = (r_1,...,r_n)\in \mathbb{R}^n$ and $\vec{c} = (c_1,...,c_n)\in \mathbb{C}^n$. $(\vec{r},\vec{c})$ is called the {\it conformal module} of $D_S$. Solving the above problem (\ref{originalproblem1}) and (\ref{originalproblem2}) is challenging since both the conformal module of the target parameter domain and the quasi-conformal map $f$ are unknown.

In this work, we propose a variational approach to solve the problem. Let $\partial S =\{\gamma_0,\gamma_1,...,\gamma_n\}$ where $\gamma_0$ is the outermost boundary, $\gamma_1,...,\gamma_n$ are the inner boundaries. Equation (\ref{originalproblem2}) is equivalent to solving:
\begin{equation}\nonumber
f = \mathbf{argmin}_{f:S\to D_S}\{||\frac{\partial f}{\partial \bar{z}}- \mu\frac{\partial f}{\partial z}||_{\infty}\}.
\end{equation}

Hence, we can set up our problem as minimizing:
\begin{equation}\label{optimization1}
E_B (f,\vec{r},\vec{c}) = \{\int_S |\frac{\partial f}{\partial \bar{z}}- \mu\frac{\partial f}{\partial z}|^p dS\},
\end{equation}
\noindent subject to the constraints that:

\noindent (1) $f|_{\gamma_i}(\gamma_i) = \partial B_{r_i}(c_i)$ for $i=1,2,...,n$ ,

\noindent (2) $f|_{\gamma_0}(\gamma_0) = \partial \mathbb{D}$ and

\noindent (3) $||\mu(f)||_{\infty} := ||\frac{\partial f}{\partial \bar{z}}/\frac{\partial f}{\partial z}||_{\infty}<1$.

The optimal map is called the quasi-conformal parameterization (QCMC) for the multiply-connected domain. Note that if we set $p$ to be large enough, the optimization problem gives a good approximation of the quasi-conformal parameterization solving equation (\ref{originalproblem2}). In practice, we set $p=2$ and it is found that the obtained optimal map is already a close approximation of the desired quasi-conformal parameterization. In this case, the optimal map is a {\it least square quasi-conformal parameterization} (LSQCMC) for $S$.

Furthermore, constraint (3) guarantees the map $f$ is bijective. This can be explained by the following theorem.

\medskip

\begin{thm}
If $f:S_1 \rightarrow S_2$ is a $C^1$ map satisfying the constraint (3) , then f is bijective .
\end{thm}

\begin{proof}
Suppose $f=u+iv$ under some local coordinates. The Beltrami coefficient $\mu_f$ is given by : $\mu(f) = \frac{\partial f}{\partial\overline{z}}/\frac{\partial f}{\partial z}$, where : $\frac{\partial f}{\partial\overline{z}}=\frac{(u_x-v_y)+i(u_y+v_x)}{2}$ and $\frac{\partial f}{\partial z}= \frac{(u_x+v_y)+i(v_x+u_y)}{2}$.

Also , the Jacobian of f , $J_f$ is given by:
\begin{equation}
\begin{split}
J_f &= u_xv_y-u_yv_x \\
&=\frac{(u_x+v_y)^2+(v_x+u_y)^2-(u_x-v_y)^2-(u_y+v_x)^2}{4} \\
&=\left|\frac{\partial f}{\partial\overline{z}}\right|^2-\left|\frac{\partial f}{\partial z}\right|^2=\left|\frac{\partial f}{\partial\overline{z}}\right|^2(1-|\mu_f|^2).
\end{split}
\end{equation}

Since $||\mu(f)||_\infty<1$ , we have $|J_f|>0$ everywhere , hence bijective .
\end{proof}

\medskip

The optimization problem (\ref{optimization1}) is not a standard $L^p$-minimization problem. It involves both the optimizations of the quasi-conformal map $f$ together with the boundary constraints. In other words, we simultaneously look for the best conformal module for the boundary constraints and the optimal quasi-conformal map $f$ satisfying the constraints such that the energy $E_B$ is minimized. Theoretically, there exists a conformal module and quasi-conformal map $f$ such that $E_B=0$.

In this paper, we propose to solve this optimization problem (\ref{optimization1}) by an iterative descent algorithm, which considers both the map $f$ and the conformal module $(\vec{r},\vec{c})$ as variables, to minimize $E_B$. We will explain the proposed iterative method in details in the following subsections.

\subsection{Energy minimization with fixed conformal module}
In this subsection, we discuss how we can iteratively adjust the map $f$ to minimize $E_B$ with fixed conformal module. More specifically, given a punctured disk $D_S = \mathbb{D}\setminus \bigcup_{i=1}^n B_{r_i}(c_i)$, we look for an optimal map $f:S\to D_S$ that minimizes $E_B$. To solve this problem, we propose to iteratively find a sequence of maps from an initial map, which converges to our desired quasi-conformal parameterization minimizing $E_B$. Let $f^{\nu}$ be an initial quasi-conformal map with Beltrami coefficient $\nu$. Our goal is to deform $f^{\nu}$ to another map $f$ such that it reduces $E_B(f)$. Let $\delta f = f - f^{\nu}$ and let $\mathcal{A}(\mu)$ be the differential operator defined by $\mathcal{A}(\mu):=\frac{\partial}{\partial\overline{z}}-\mu\frac{\partial}{\partial z}$. Then, the energy functional $E_B(f)$ can be reformulated as follows:
\begin{equation}\label{variatin}
E_B(f) = E_B'(\delta f) := \int_S \left\vert\mathcal{A}(\mu)\delta f + \mathcal{A}(\mu)f^{\nu} \right \vert^p dS.
\end{equation}

Using the above formualtion, we propose to iteratively deform an initial map $g^0:S\to D_S$ to an optimal quasi-conformal map $g^*:S\to D_S$, whose Beltrami coefficient is closest to the given $\mu$ in the $L^p$-sense. The initial map can be chosen as the harmonic map with boundary pointwise correspondence given by the arc-length parameterization. Let the Beltrami coefficient of $g^0$ be $\nu_0$. Our goal is to deform $g^0$ to $g^1= g^0 + \delta g^0$, whose Beltrami coefficient is closer to the target $\mu$ in the $L^p$-sense. According to new formulation (\ref{variatin}), it can be achieved by finding a $\delta g^0$ that minimizes
\begin{equation}\label{vectorenergy}
E_B'(\delta g^0)= \int_{S} |\mathcal{A}(\mu)\delta g^0 + \mathcal{A}(\mu)g^0|^p dS
\end{equation}
\noindent subject to the boundary constraints: $g^1|_{\gamma_i}(\gamma_i) = \partial B_{r_i}(c_i)$ and $g^1|_{\gamma_0}(\gamma_0) = \partial \mathbb{D}$.

The boundary constraints can be enforced as follows. In general, points on the boundary $\gamma_i$ of $S$ are allowed to move along the target boundary $\gamma_i' \subset \partial D_S$ ($\gamma_i' = \partial B_{r_i}(c_i)$ if $i=1,2,...,n$ and $\gamma_0' = \partial \mathbb{D}$). In other words, we need to restrict the movement to be the tangential direction of the target boundary component. Let $u_k\in \gamma_i$ be a point on the boundary component $\gamma_i$ of $S$. We require that the variation $\delta g^0$ on $u_k$ satisfies the following:
\begin{equation}\label{tangential}
\delta g^0(u_k) = t_k \mathbf{T}_{\gamma'_k}(g^0(u_k)),
\end{equation}
\noindent where $\mathbf{T}_{\gamma'_k}(g^0(u_k))$ is the unit tangent vector of $\gamma'_k$ at $g^0(u_k)$ and $t_k\in \mathbb{R}$ is a variable measuring the length of $\delta g^0$ at $u_k$ (how far it deforms the map $g^0$).

Note that the target boundary component $\gamma'_k$ is a circle. Suppose $\gamma'_k$ is a circle centered at $c_k$ with radius $r_k$, the equation can be simplified as:
\begin{equation}\label{tangential2}
\delta g^0(u_k) =it_k \frac{g^0(u_k)-c_k}{r_k}.
\end{equation}

With this setting, our problem becomes solving the optimization problem (\ref{vectorenergy}) subject to the boundary constraint (\ref{tangential2}). In practice, we choose $p=2$. In this case, the problem becomes a least square minimization problem. In the discrete case, it is equivalent to a sparse linear system.

Now, as the boundary point $u_k$ moves along the tangential direction, it may leave the target boundary $\gamma'_k$ by a short distance. Thus, we project it back to $\gamma'_k$ by solving:
\begin{equation}\label{projection}
g^1(u_k) = \mathbf{argmin}_{x \in \gamma'_k}\left\Vert g^0(u_k)+\delta g^0(u_k)-x \right\Vert_2 ^2
\end{equation}
Since $\gamma'_i$ is simply a circle, this process is just rescaling the distance between $g^0(u_k)+\delta g^0(u_k)$ and $c_k$ to $r_k$ as follows:
\begin{equation}\label{projection2}
g^1(u_k) = \frac{g^0(u_k)+\delta g^0(u_k)-c_k}{\left\Vert g^0(u_k)+\delta g^0(u_k)-c_k \right \Vert} \cdot r_k +c_k.
\end{equation}

Note that, according to our setting, we only allow the boundary points to move a bit along the target boundary. Although $g^1 := g^0 + \delta g^0$ reduces the energy $E_{B}'$, it is unlikely the optimal map that solves the optimization problem (\ref{vectorenergy}) and (\ref{tangential2}).

Therefore, we repeat the procedure. More precisely, suppose $g^k$ is obtained at the $k$-th iteration. We can obtain a new map $g^{k+1}:= g^k+\delta g^k$ that reduces $E_{B}'$ by minimizing $||\mathcal{A}(\mu)\delta g^k + \mathcal{A}(\mu) g^k||_p^p$ subject to the boundary constraint (\ref{tangential2}). As a result, we obtain a sequence of maps $\{g^k\}_{k=1}^{\infty}$ with Beltrami coefficients $\{\mu^k\}_{k=1}^{\infty}$, which converges to the desired optimal map.

In summary, the iterative scheme for computing the optimal quasi-conformal map with fixed conformal module of the target parameter domain can be described as Algorithm \ref{algorithm1}.

\begin{algorithm}\label{algorithm1}
 \KwData{triangular mesh of the source domain $S$, target domain $D_S$ and target Beltrami coefficient $\mu$.}
 \KwResult{$g^*:S_1 \rightarrow D_S$}.
Initialize $g^0$ to be harmonic.

Calculate the Beltrami coefficient $\mu^0$ of $f^0$.

 \While{$\left\Vert \mu_n - \mu_{n-1} \right\Vert > \epsilon$}{
  Find $\delta g^{n-1}$ that optimizes (\ref{vectorenergy}) subject to the constraint (\ref{projection2}).

  Set $g^n=g^{n-1}+\delta g^{n-1}$.
}
 \caption{Energy minimization for fixed conformal module}
\end{algorithm}

\subsection{Adjustment of the conformal module}
One of the main challenge for solving the optimization problem (\ref{optimization1}) is that the boundary constraints is not fixed. In other words, the geometry of the target parameter domain or the conformal module can be adjusted to further minimize $E_B'$.

To solve this problem, our strategy is to regard the conformal module as another variable, and let it vary together with the map during the optimization process. In other words, we need to incorporate the conformal module into the energy functional of the optimization problem.

Given a multiply-connected domain $S$, we first obtain an intial guess of the conformal module $(\mathbf{r}^0,\mathbf{c}^0)$. $\mathbf{r}^0 = (r_0,r_1,...,r_n)^T$ and $\mathbf{c}^0=(c_0,c_1,...,c_n)^T$ can be chosen as follows. For each $\gamma_i$ ($i=0,1,2,...,n$), we find a circle approximating $\gamma_i$. In this way, we obtain a circle domain (a disk with $n$ inner disks removed). The circle domain is then normalized such that the outermost boundary becomes the unit circle. The initial guess $\mathbf{r}^0$ and $\mathbf{c}^0$ are then obtained, and the initial guess of the parameter domain $D_S^0$ is obtained. In other words, $D_S^0 = \mathbb{D}\setminus \bigcup_{i=1}^n B_{r_i^0}(c_i^0)$.

Our problem can now be regarded as finding a sequence of punctured disks $D_S^i$ and maps $g^i$, which iteratively minimizes $E_B(g,\mathbf{r},\mathbf{c})$. To do this, we propose to give an extra freedom to the variation $\delta g^i$ on the boundary component $\gamma_k$ ($k=1,2,...,n$). The main idea is to allow $\mathbf{r}^i$ and $\mathbf{c}^i$ to vary. More specifically, the translation of the center from $c_k^i$ to $c_k^i + \Delta c_k^i$ can be considered as all points on $\gamma_k$ being translated by $\Delta c_k^i$. Also, the scaling of the radius from $r_k^i$ to $r_k^i + \Delta r_k^i$ can be regarded as all points on $\gamma_k$ being moved along the radial direction by $\Delta r_k^i$. Therefore, the variation $\delta g^i$ on the boundary point $u_k\in \gamma_k$ can be formulated as:
\begin{equation}\label{tangential3}
\delta g^i(u_k) = t_k \mathbf{T}_{\gamma'_k}(g^i(u_k))+\Delta c_k^i+\frac{u_k-c_k^i}{\left\Vert u_k-c_k^i \right\Vert} \Delta r_k^i.
\end{equation}

Note that the target boundary of the outermost boundary is always fixed to be the unit circle. Thus, the above formulation does not apply to $\gamma_0$.

We then minimize the energy functional (\ref{vectorenergy}) subject to the boundary constraint (4.8) and (\ref{tangential3}). This new optimization problem involves new variables ${\Delta \mathbf{c}}^i$ and ${\Delta \mathbf{r}}^i$, which are the perturbation of the centers and radii (conformal module) of the inner disks of $D_S^i$. With this setting, we simultaneously optimize the map $g^i$ and the conformal module $(\mathbf{r}^i, \mathbf{c}^i)$ to optimize $E_B'$.

Note that the variation $\delta g^i$ for the interior points is a complex number, which can be represented by two real scalars (one for the real part and one for the imaginary part). Besides, the variation $\delta g^i$ for the boundary points must be a scalar multiple of the tangential direction of the boundary and can be represented by one real scalar. Denote the representation of $\delta g$ by $\widetilde{\delta g}^i$. A linear operator $K^i$ transforms the representation $\widetilde{\delta g}^i$ back to $\delta g^i$. $K^i$ depends on the variables $\Delta \mathbf{r}^i$ and $\Delta \mathbf{c}^i$. Hence, our problem can be formulated as minimizing the following energy functional over $\widetilde{\delta g}^i$:
\begin{equation}\label{KLinear}
E_B'(K^i \widetilde{\delta g}^i) = \int_S \left\vert \mathcal{A}(\mu) K^i \widetilde{\delta g}^i + \mathcal{A}(\mu) g^i \right \vert^p dS.
\end{equation}

Representing $\delta g^i$ by its representation $\widetilde{\delta g}^i$ reduces storage requirement. More importantly, it allows us to incorporate the conformal module into the energy functional of the optimization problem. In practice, we solve (\ref{KLinear}) by taking $p=2$. In other words, we solve (\ref{KLinear}) by the least square method.

The overall iteratve scheme is summarized in Algorithm \ref{algorithm2}.

\begin{algorithm}\label{algorithm2}
 \KwData{triangular mesh of source domain $S$, target Beltrami coefficient $\mu$, tolerance $\epsilon$.}
 \KwResult{Quasi-conformal parameterization $g^*:S\rightarrow D_S$.}
 Initialize $D_0$ to be punctured disk with the same topology of $S$.

Initializa $g^0 : \Omega\rightarrow D_0$ to be the harmonic map.

Calculate the Beltrami coefficient $\mu^0$ of $g^0$.

 \While{$\left\Vert \mu_{n+1} - \mu_n \right\Vert > \epsilon$}{
Calculate tangent vector and outward normal vector at boundary.

Construct constraint matrix $K^n$.

Find $\delta g^n$, $\Delta \mathbf{c}^n$ and $\Delta \mathbf{r}^n$ by solving (\ref{KLinear}) with $L^p$-minimization.

Let $g^{n+1} = g^n+t \cdot K^n\delta g^n$ (e.g. take $t=0.5$).

Update  $g^{n+1}(u_k) \leftarrow \frac{g^n(u_k)-c_k^n}{\left\Vert g^n(u_k)-c_k^n \right \Vert} \cdot r_k^n +c_k^n$ for all $u_k\in \gamma_k$.

}
 \caption{Overall algorithm of QCMC}
\end{algorithm}

\section{Numerical implementation}\label{implementation}
In this section, we will describe briefly how the proposed algorithms introduced in Section \ref{proposed} can be implemented. The major components of the proposed algorithms are the constrution of $\mathcal{A}$ and $K^n$ in the discrete setting. In the following subsection, we will describe how to discretize the two operators.

In practice, 2D domains or surfaces in $\mathbb{R}^3$ are usually represented discretely by triangular meshes. Suppose $M$ is the surface mesh representing the surface $S$. We define the set of vertices on $M$ by $V = \{v_i\}_{i=1}^n$. Similarly, we define the set of triangular faces on $M$ by $F = \{T_j\}_{j=1}^m$. Our goal is to look for a quasi-conformal parameterization $\vec{f}:M\to \mathbb{C}$. We further introduce the following notions.

\medskip

\noindent \begin{itemize}
\item $U$ is a $2n \times 1$ vector. It stores the position of $\{\vec{f}(v_i)\}_{i=1}^n$, namely, ${\bf U}=(Re(\vec{f}(v_1)), Re(\vec{f}(v_2)),...,Re(\vec{f}(v_n)), Im(\vec{f}(v_1)), Im(\vec{f}(v_2)),...,Im(\vec{f}(v_n)))^T$.

\item $V_n$ is a $2n \times 1$ vector. It stores the vector field acting on ${\bf U}$ at the $n^{th}$ iteration.

\item $A$ is a $2m\times 2m$ matrix, which is the matrix representation of the differential operator: $\mathcal{A} = \frac{\partial}{\partial\overline{z}}-\mu\frac{\partial}{\partial z}$ .

\end{itemize}

\subsection{Construction of the matrix representation $A$ of $\mathcal{A}$}

In practice, we consider the parameterization $\vec{f}:M\rightarrow D$ to be piecewise linear, where $D$ is the triangulation mesh of a punctured disk. In other words, we regard the restriction of $\vec{f}$ on each trianguar face $T_i$ as an affine transform. Hence,
$$f_i:=f|_{T_i}=\frac{\partial f_i}{\partial z}z+\frac{\partial f_i}{\partial\overline{z}}\overline{z}+\delta_i.$$

\noindent Let the vertices of $T_i$ be $\{w_1 , w_2 , w_3\}$ and vertices of $f_i(T_i)$ be $\{z_1,z_2,z_3\}$. Then, $\frac{\partial f_i}{\partial z}$ and $\frac{\partial f_i}{\partial\overline{z}}$ satisfy the following linear system :

$$\left( \begin{array}{ccc}w_1 & \overline{w_1} & 1 \\
w_2 & \overline{w_2} & 1 \\
w_3 & \overline{w_3} & 1 \end{array} \right)
\left( \begin{array}{c} \frac{\partial f_i}{\partial z} \\ \frac{\partial f_i}{\partial\overline{z}} \\ \delta_i \end{array}\right)
= \left( \begin{array}{c} z_1 \\ z_2 \\ z_3 \end{array} \right).
$$

\noindent The above linear system can be reduce to:

$$\left( \begin{array}{ccc}w_2-w_1 & \overline{w_2-w_1}  \\
w_3-w_1 & \overline{w_3-w_1}\end{array} \right)
\left( \begin{array}{ccc} \frac{\partial f_i}{\partial z} \\ \frac{\partial f_i}{\partial\overline{z}} \end{array}\right)
= \left( \begin{array}{ccc} z_2-z_1 \\ z_3-z_1\end{array} \right),$$

Hence,

$$
\left( \begin{array}{ccc} \frac{\partial f_i}{\partial z} \\ \frac{\partial f_i}{\partial\overline{z}} \end{array}\right)
= \left( \begin{array}{ccc}w_2-w_1 & \overline{w_2-w_1}  \\
w_3-w_1 & \overline{w_3-w_1}\end{array} \right)^{-1} \left( \begin{array}{ccc} z_2-z_1 \\ z_3-z_1\end{array} \right).$$

\noindent By solving this linear system, we get $\frac{\partial f_i}{\partial z}$ and $\frac{\partial f_i}{\partial\overline{z}}$. With that, $\mathcal{A}f_i$ can be constructed by $\mathcal{A}f_i=\frac{\partial f_i}{\partial\overline{z}}-\mu(T_i)\frac{\partial f_i}{\partial z}$.

Also, $(Re(\mathcal{A}f_1),Re(\mathcal{A}f_2),...,Re(\mathcal{A}f_m), Im(\mathcal{A}f_1),Im(\mathcal{A}f_2),...,Im(\mathcal{A}f_m))^T$ can be written as $A U$, where $A\in M_{2m\times 2m}(\mathbb{R})$. Hence, the matrix representation $A$ of $\mathcal{A}$ can be obtained.

\subsection{Construction of the boundary constraint matrix $K$}
Now, we discuss how to construct the boundary constraint matrix $K$.

Let $I_{int} =\{i_1,i_2,...,i_r\}$ be the indices of all interior vertices.

Let  $I_{bdy} = \{i_1^b,...,i_s^b\}$ be the indices of all boundary vertices.

Denote the tangent vector of the boundary vertex $v_{i_k^b}$ by $t_k\in \mathbb{C}$.

Note that $r+s = n$. Under the boundary constraints, the admissible variation $\delta g$ can be represented by a vector $\widetilde{\delta g} \in \mathbb{R}^{n+r}$. In particular, the variation on the interior vertices depends on two scalar variabes (real and imaginary parts of the vector field). Hence, $2r$ scalars are required to represent the variation on the interior vertices. For the boundary vertices, the variation must be a scalar multiple of the tangential direction and it depends on one scalar variable. Thus, $s$ scalars are required to represent the variation on the boundary vertices. All together, $2r+ s = n+r$ scalars are required to represent the admissible variation.

The boundary constraint matrix $K$ transforms the variation representation $\widetilde{\delta g}$ into the admissible variation $\delta g$. The matrix $K$ can be regarded as a $2n\times (n+r)$ sparse matrix, which can be constructed as follow :

For $j=1,2,...,r$ ,
\begin{equation}
\begin{array}{lll}
K_{i,j} & = & 1\\
K_{n+i,j} & = & 1
\end{array}
\end{equation}

For $j=1,2,...,s$ ,
\begin{equation}
\begin{array}{lll}
K_{i^b_j,r+j} & = & Re(t_j)  \\
K_{n+i^b_j,r+j} & = & Im(t_j)
\end{array}
\end{equation}

In this way, $\delta g :=K\widetilde{\delta g} \in \mathbb{R}^{2n}$ becomes a vector of size $2n$. The first $n$ entries represent the real part of the variation and the last $n$ entries represent the imaginary part of the variation.

\section{Experimental results}\label{experiment}
In this section, we test our proposed algorithm to compute conformal and quasi-conformal parameterizations of synthetic 2D multiply-connected domains and multiply-connected real human face surfaces. All our experiments are carried out on a desktop with following specification:

\begin{table}[!h]
\begin{center}
\begin{tabular}{c||c}
 Hardware & Specification  \\
\hline
Processor & AMD A8-6600K (3.9GHz $\times 4$) \\
Memory & 8GB DDR3-1600 \\
Operating system & Window 8 Home Pre. 64bit \\
Experiment perform & Matlab 2013a \\
GPU assisted computation & No \\
\end{tabular}
\label{table:computer specification}
\end{center}
\end{table}

\subsection{Synthetic 2D domains}
We first examine our algorithm on multiply-connected 2D domains.

\begin{figure}[h]
\centering
\includegraphics[width=3.75in]{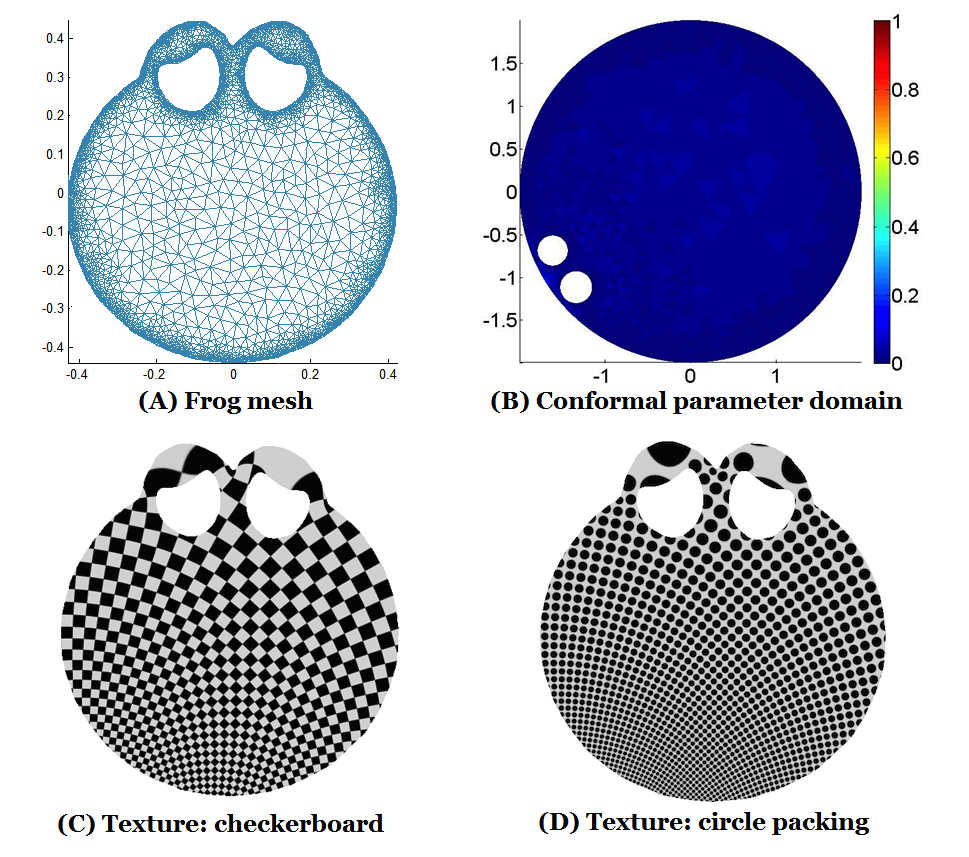}
\caption{(A) shows the frog mesh. (B) shows the conformal parameterization. (C) shows the texture mapping of the checkerboard using the obtained conformal parameterization. (D) shows the texture mapping of the circle packing pattern.\label{fig:frogexample1}}
\end{figure}

\begin{figure}[h]
\centering
\includegraphics[width=3.75in]{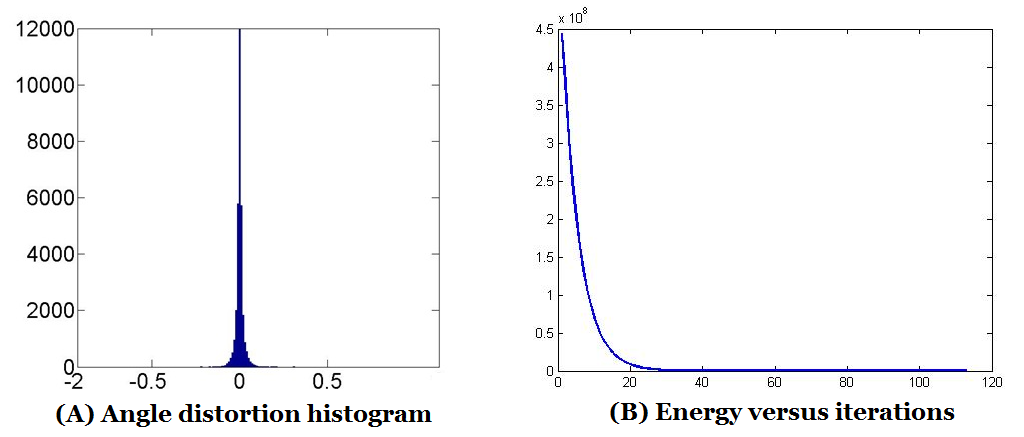}
\caption{(A) shows the histogram of the angle distortion under the conformal parameterization of the frog mesh. (B) shows the energy versus iterations.\label{fig:frogexample2}}
\end{figure}

\paragraph{Example 1} In this example, we compute the least square conformal parameterization of a 2D triply-connected frog mesh, which is shown in Figure \ref{fig:frogexample1}(A). The conformal parameterization is shown in (B). The colormap is given by the norm of the Beltrami coefficient. The blue color indictates the Beltrami coefficient is close to 0, meaning that the parameterization is indeed conformal. Using the conformal parameterization, we map the checkerboard texture and circle packing texture onto the frog mesh, as shown in (C) and (D). Note that the right angle of the checkerboard pattern are well-preserved, meaning that the parameterization is angle-preserving. The circle pattern of the circle packing texture is also preserved. It means that under the conformal parameterization, infinitesimal circles are mapped to infinitesimal circles as expected. Figure \ref{fig:frogexample2}(A) shows the histogram of the angle distortion under the parameterization. It accumultes at 0, meaning that the parameterization is indeed angle-preserving. The energy versus iterations using our proposed algorithm is shown in (B). $E_{B}$ decreases as iteration increases.

\medskip

\begin{figure}[h]
\centering
\includegraphics[width=3.75in]{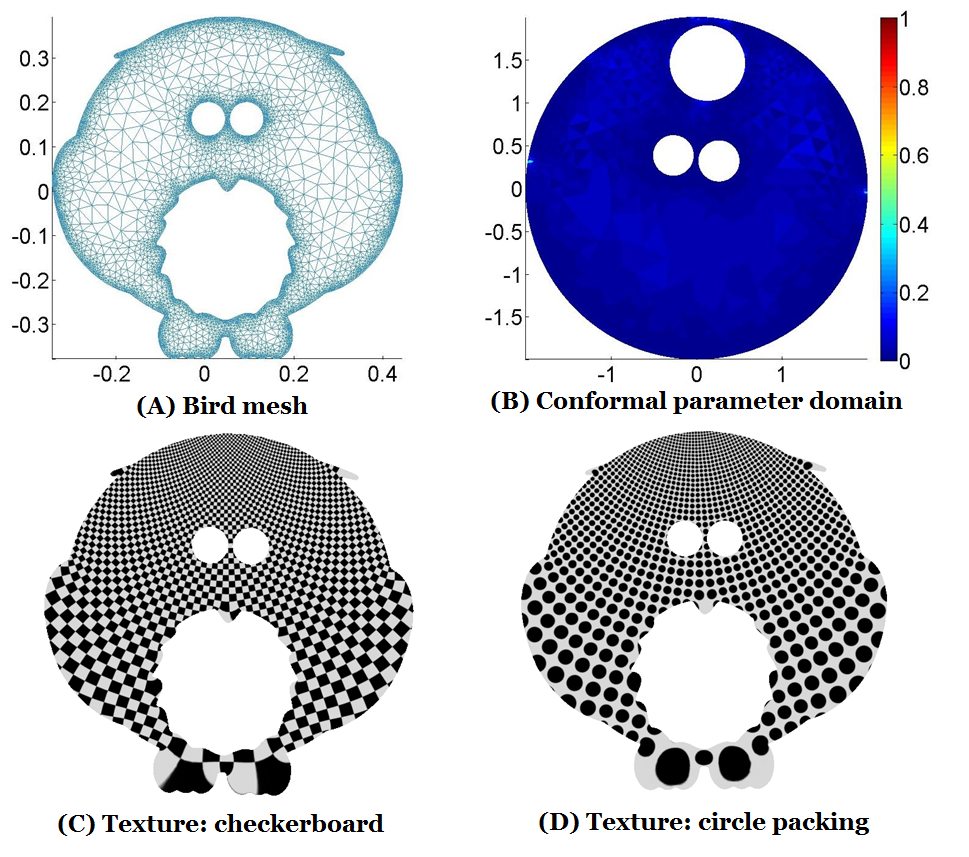}
\caption{(A) shows the bird mesh. (B) shows the conformal parameterization. (C) shows the texture mapping of the checkerboard using the obtained conformal parameterization. (D) shows the texture mapping of the circle packing pattern.\label{fig:birdexample1}}
\end{figure}

\begin{figure}[h]
\centering
\includegraphics[width=3.75in]{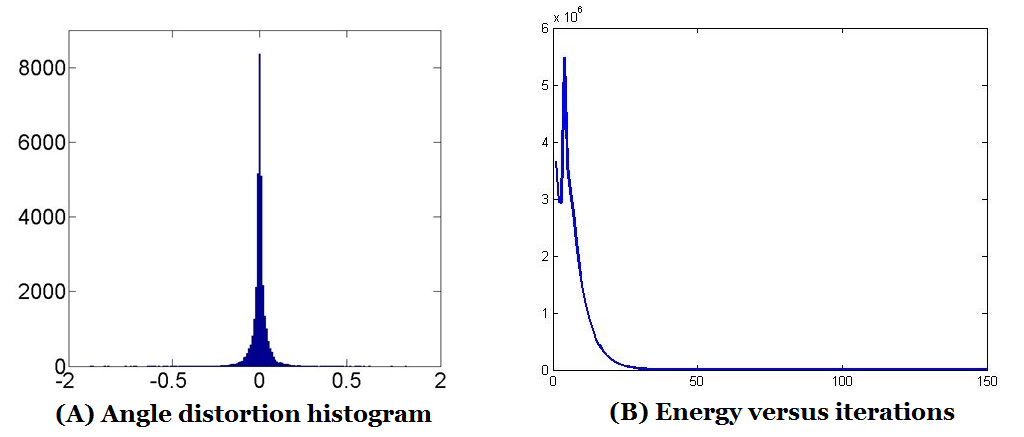}
\caption{(A) shows the histogram of the angle distortion under the conformal parameterization of the bird mesh. (B) shows the energy versus iterations.\label{fig:birdexample2}}
\end{figure}

\paragraph{Example 2} We next test the proposed algorithm on a 2D bird mesh, which is shown in Figure \ref{fig:birdexample1}(A). The conformal parameterization is shown in (B), whose colormap is given by the norm of the Beltrami coefficient. The texture map of the checkerboard and circle-packing pattern are shown in (C) and (D) respectively, indicating the parameterization is angle-preserving. Figure \ref{fig:birdexample2}(A) shows the histogram of the angle distortion under the parameterization. It accumultes at 0, meaning that the parameterization is indeed angle-preserving. The energy versus iterations using our proposed algorithm is shown in (B).

\medskip

\begin{figure}[h]
\centering
\includegraphics[width=3.75in]{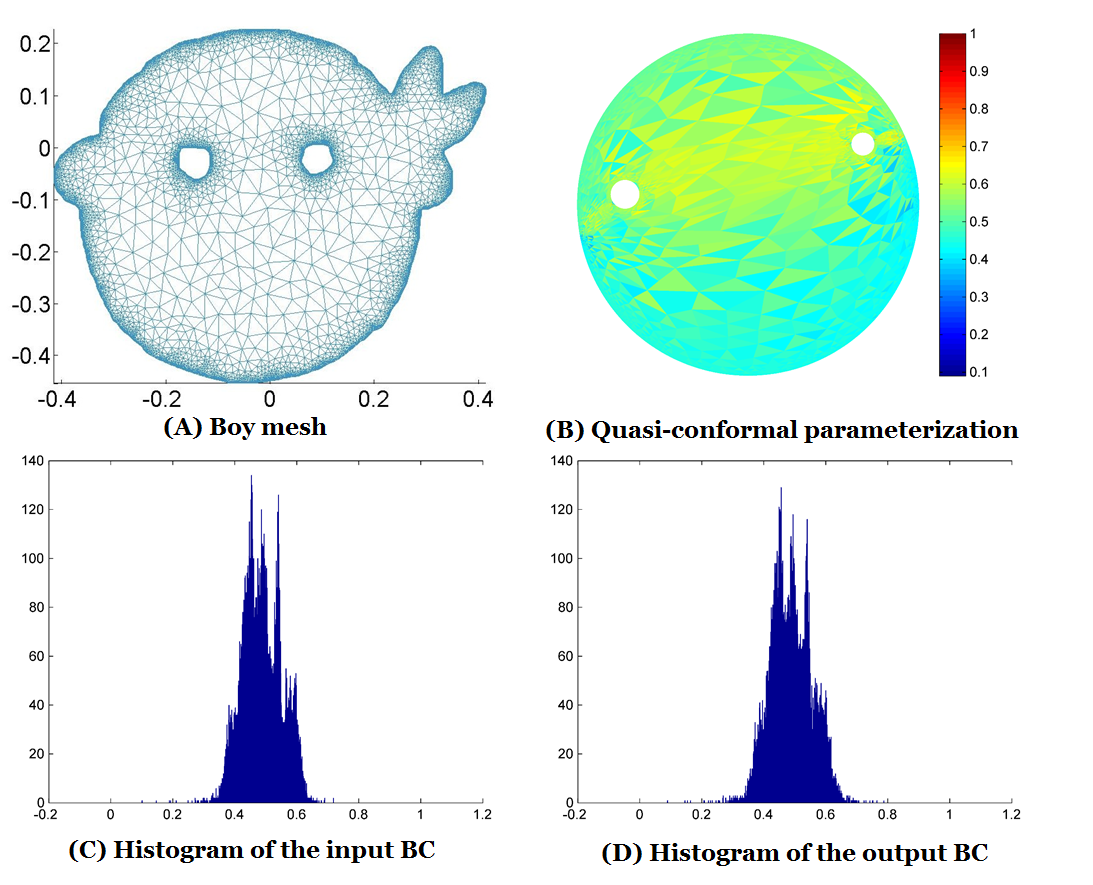}
\caption{(A) shows the boy mesh. (B) shows the quasi-conformal parameterization. (C) shows the histogram of the target Beltrami coefficient. (D) shows the histogram of the norm of the output Beltrami coefficient.\label{fig:boyexampleqc1}}
\end{figure}

\begin{figure}[h]
\centering
\includegraphics[width=3.75in]{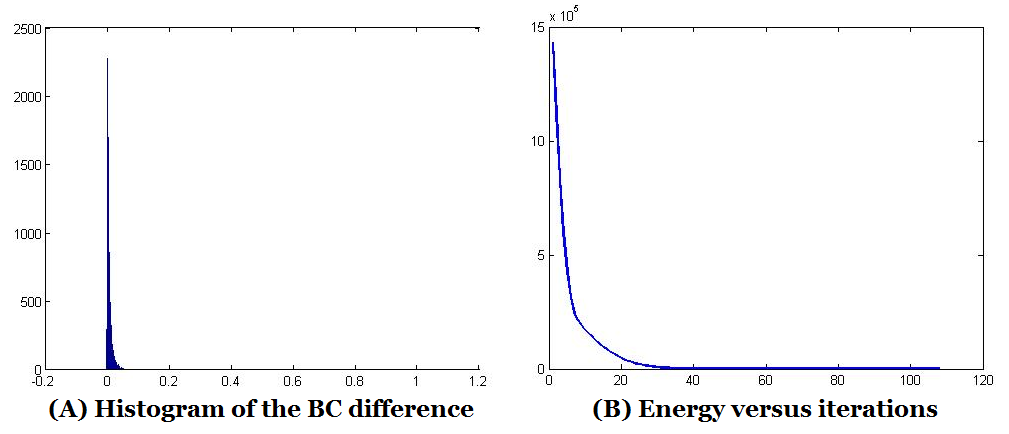}
\caption{(A) the histogram of the error between the target and output Beltrami coefficients for the boy mesh example. (B) shows the energy versus iterations.\label{fig:boyexampleqc2}}
\end{figure}

\paragraph{Example 3} We test our algorithm to compute the least square quasi-conformal parameterization of a 2D triply-connected boy mesh, which is shown in Figure \ref{fig:boyexampleqc1}(A). The quasi-conformal parameterization is shown in (B). (C) shows the histogram of the norm of the target Beltrami coefficient. (D) shows the histogram of the norm of the output Beltrami coefficient. It closely resemble to that of the target Beltrami coefficient. It means that the Beltrami coefficient of the quasi-conformal parameterization is close to our target Beltrami coefficient. Figure \ref{fig:boyexampleqc2} shows the histogram of the error between the target and output Beltrami coefficients. It accumulates at 0, indicating that the obtained map is a good approximation to our desired quasi-conformal parameterization. The energy versus iterations using our proposed algorithm is shown in (B).

\begin{figure}[h]
\centering
\includegraphics[width=3.75in]{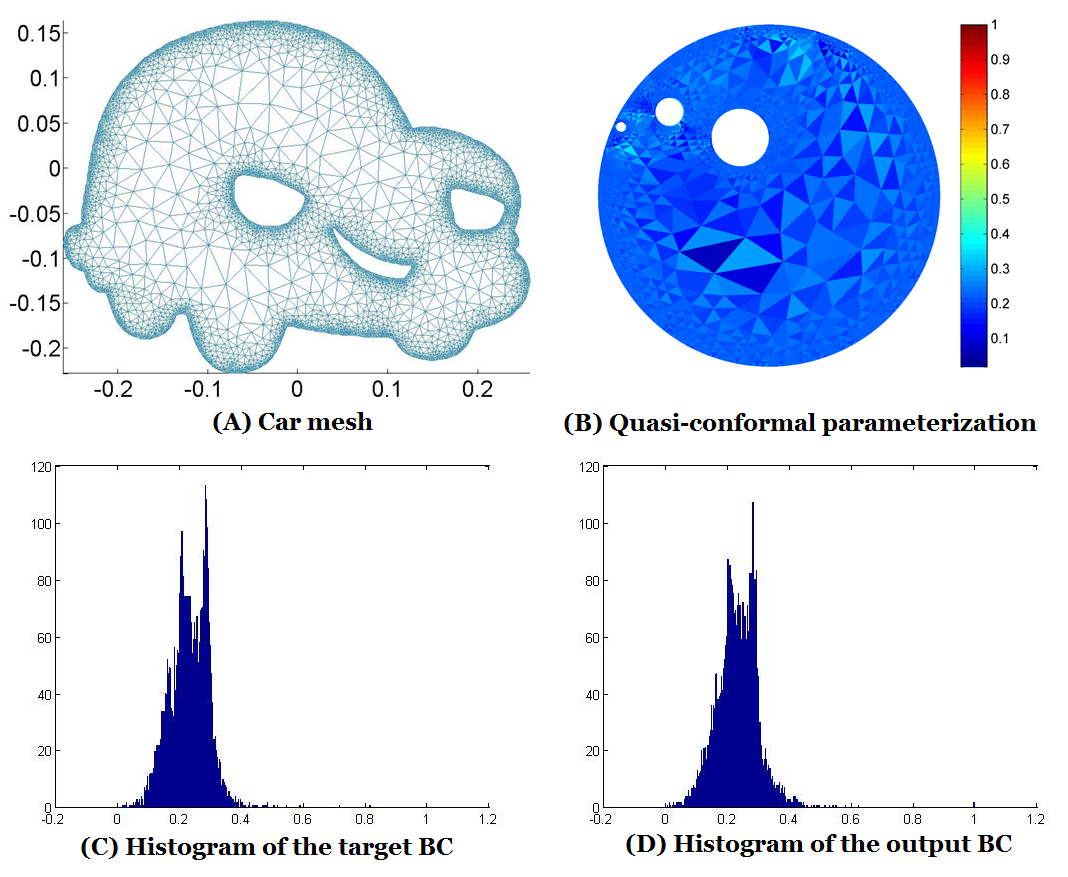}
\caption{(A) shows the car mesh. (B) shows the quasi-conformal parameterization. (C) shows the histogram of the target Beltrami coefficient. (D) shows the histogram of the norm of the output Beltrami coefficient.\label{fig:carexampleqc1}}
\end{figure}

\begin{figure}[h]
\centering
\includegraphics[width=3.75in]{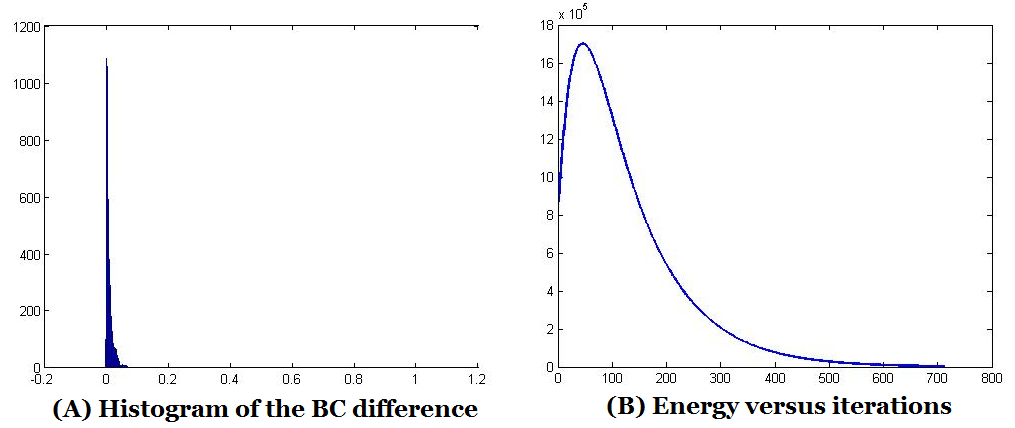}
\caption{(A) shows the histogram of the error between the target and output Beltrami coefficients for the car mesh example.(B) shows the energy versus iterations.\label{fig:carexampleqc2}}
\end{figure}

\paragraph{Example 4} In this example, we compute the quasi-conformal parameterization of a 2D car mesh with three inner holes. The mesh is shown in Figure \ref{fig:carexampleqc1}(A). The quasi-conformal parameterization is shown in (B). The histogram of the norm of the output Beltrami coefficient is shown in (D), which closely resembles to that of the target Beltrami coefficient as shown in (C). Figure \ref{fig:carexampleqc2} shows the histogram of the error between the target and output Beltrami coefficients, which accumulates at 0, indicating that the obtained map is a good approximation to our desired quasi-conformal parameterization. The energy versus iterations is shown in (B).

\subsection{Multiply-connected Riemann surfaces}
We next examine the algorithm on multiply-connected Riemann surfaces.

\begin{figure}[h]
\centering
\includegraphics[width=3.75in]{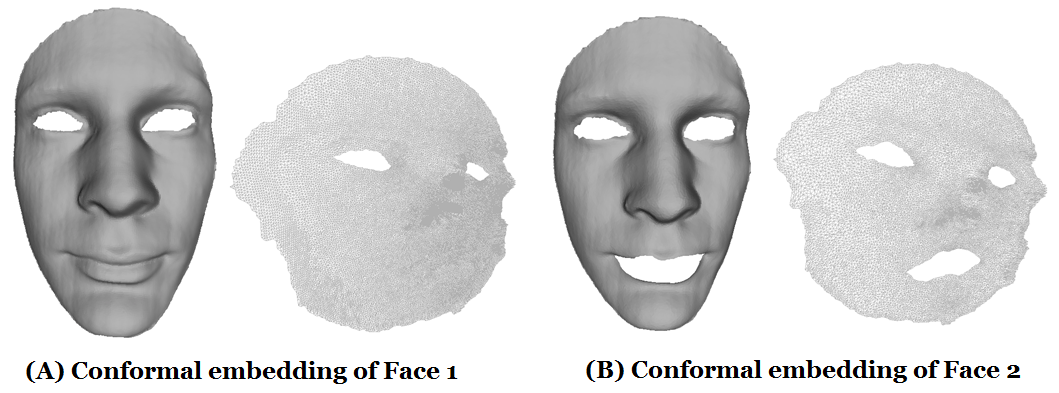}
\caption{(A) shows a triply-connected human face and its conformal embedding with free boundary condition in $\mathbb{R}^2$. (B) shows the human face with three interior regions removed and its conformal embedding with free boundary condition in $\mathbb{R}^2$. \label{fig:conformalembed}}
\end{figure}

\begin{figure}[h]
\centering
\includegraphics[width=3.75in]{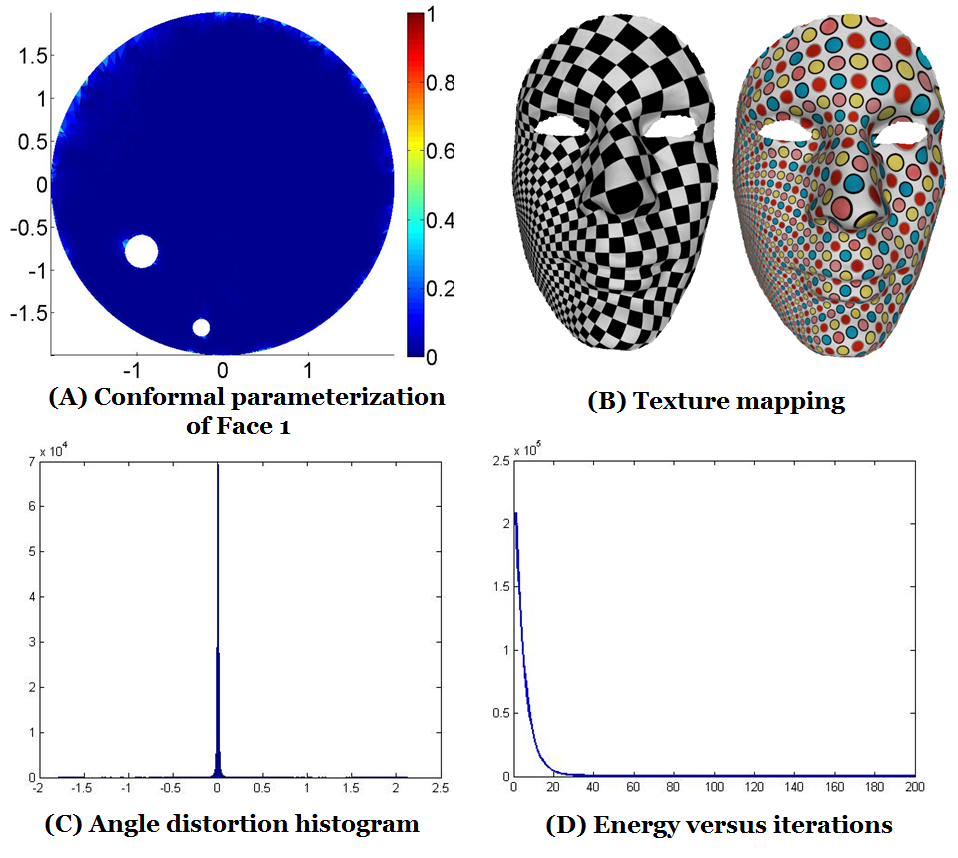}
\caption{(A) shows the conformal parameterization of human face 1. The texture map (checkerboard and circle packing) using the conformal parameterization is shown in (B). (C) shows the histogram of angle distortion. (D) shows the energy versus iterations.\label{fig:face1example1}}
\end{figure}

\paragraph{Example 5} In this example, we test the algorithm to compute the conformal parameterization of a triply-connected human face onto a punctured disk. The human face is shown in Figure \ref{fig:conformalembed}(A), and the surface is conformally embeded into $\mathbb{R}^2$ with free boundary condition. The embedded domain is then conformally parameterized onto the puntured disk. The conformal parameterization of the human face is shown in Figure \ref{fig:face1example1}(A). The texture map (checkerboard and circle packing) using the conformal parameterization is shown in (B). The right angle structure is well-preserved and the parameterization maps infinitesimal circles to circles. (C) shows the histogram of angle distortion, which shows that the parameterization is angle-preserving. (D) shows the energy versus iterations.

\begin{figure}[h]
\centering
\includegraphics[width=3.75in]{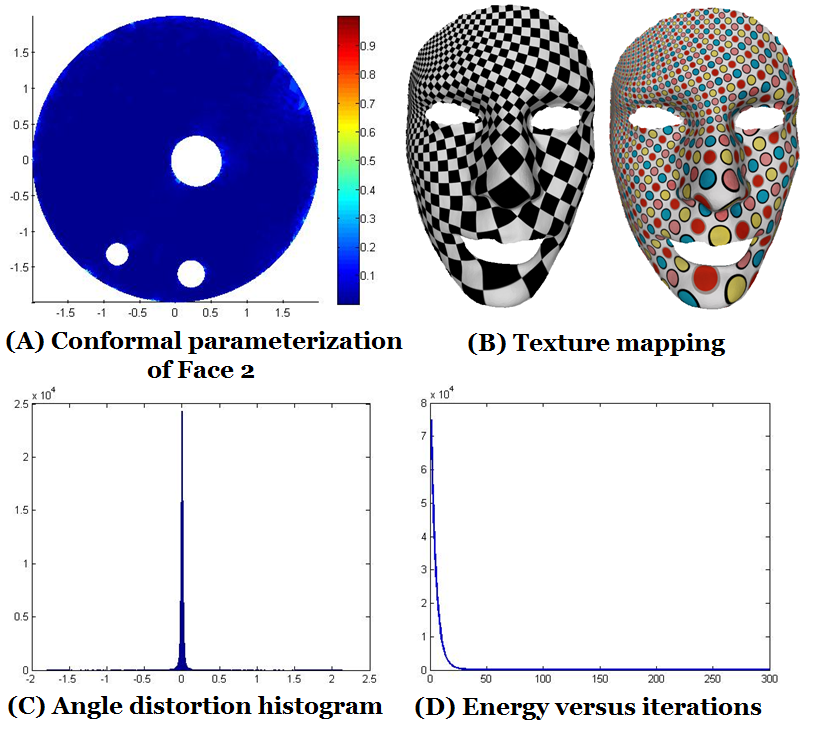}
\caption{(A) shows the conformal parameterization of human face 2. The texture map (checkerboard and circle packing) using the conformal parameterization is shown in (B). (C) shows the histogram of angle distortion. (D) shows the energy versus iterations.\label{fig:face2example1}}
\end{figure}

\paragraph{Example 6} We compute the conformal parameterization of a human face with three interior regions removed, which is shown in Figure \ref{fig:conformalembed}(B). The surface is again conformally embedded into $\mathbb{R}^2$ with free boundary conditions. Figure \ref{fig:face2example1}(A) shows the conformal parameterization. The texture maps (checkerboard and circle-packing) using the parameterization are shown in (B). The parameterization is angle-preserving, which is demonstrated from the histogram of the angle distortion in (C). (D) shows the energy versus iterations.

\begin{figure}[h]
\centering
\includegraphics[width=3.75in]{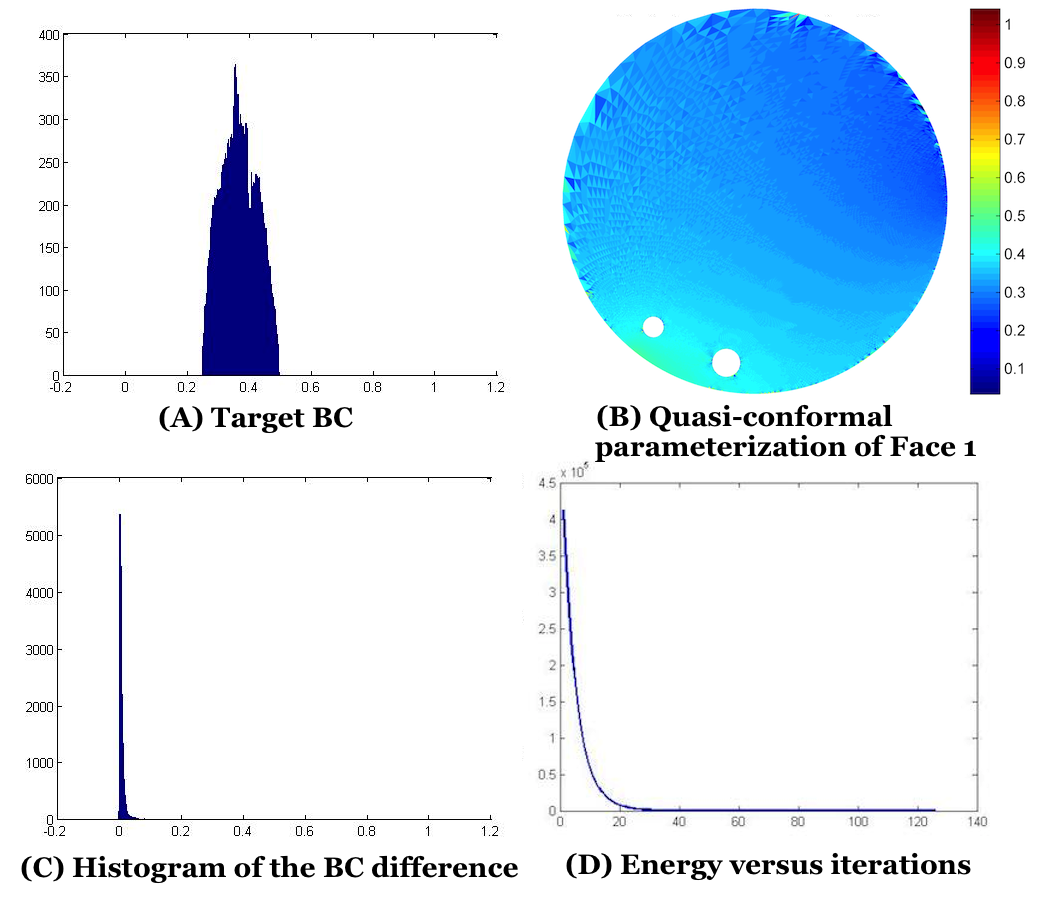}
\caption{Quasi-conformal parameterization of the human face 1. (A) shows the norm of the target Beltrami differential. The quasi-conformal parameterization is shown in (B). The colormap shows the norm of the Beltrami differential of the parameterization. (C) shows the histogram of the error between the target and output Beltrami differentials. (D) shows the energy versus iterations.
\label{fig:face1qcexample1}}
\end{figure}

\paragraph{Example 7} Now, we compute the quasi-conformal parameterization of the triply-connected human face in Figure \ref{fig:conformalembed}(A). The norm of the target Beltrami differential is shown in Figure \ref{fig:face1qcexample1}(A). The quasi-conformal parameterization is shown in (B). The colormap shows the norm of the Beltrami differential of the parameterization. (C) shows the histogram of the error between the target and output Beltrami differential. It accumulates at 0, meaning that the computed quasi-conformal parameterization is accurate. (D) shows the energy versus iterations.

\begin{figure}[h]
\centering
\includegraphics[width=3.75in]{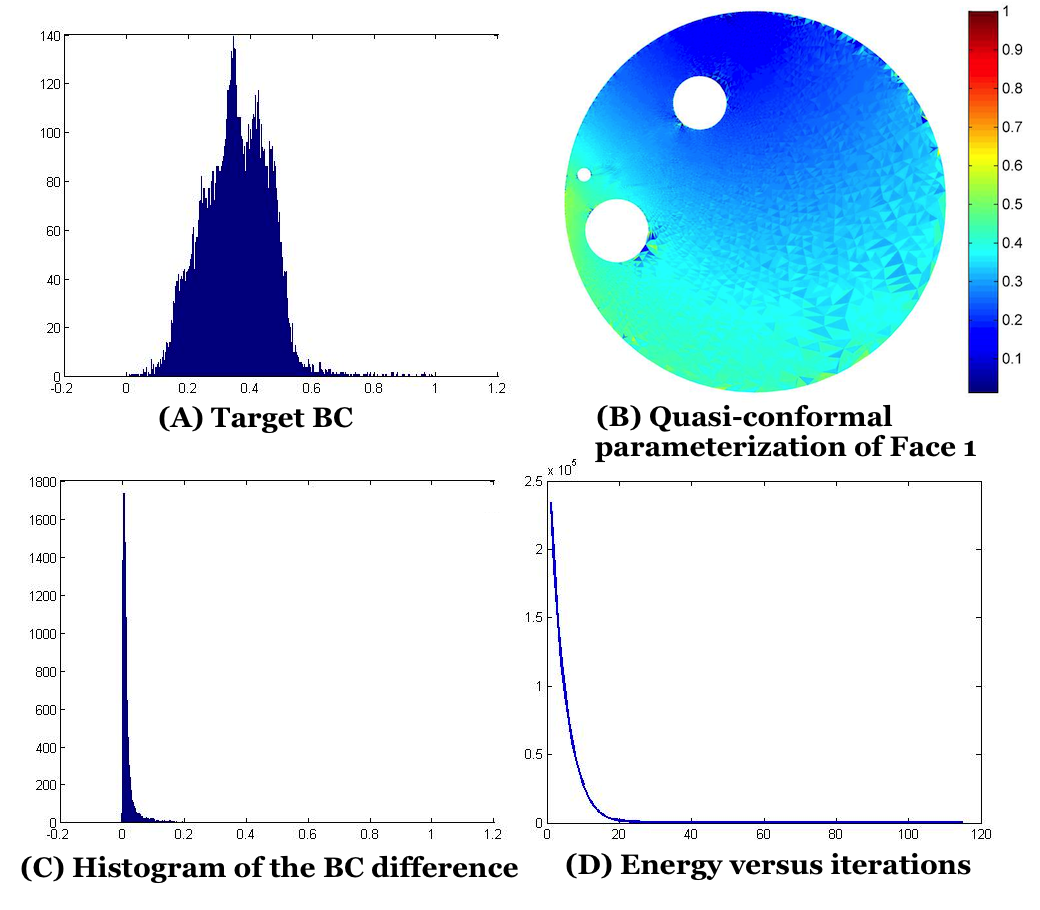}
\caption{Quasi-conformal parameterization of the human face 2. (A) shows the norm of the target Beltrami differential. The quasi-conformal parameterization is shown in (B). The colormap shows the norm of the Beltrami differential of the parameterization. (C) shows the histogram of the error between the target and output Beltrami differentials. (D) shows the energy versus iterations.
\label{fig:face2qcexample1}}
\end{figure}

\paragraph{Example 8} We also compute the quasi-conformal parameterization of the human faces with three interior regions removed in Figure \ref{fig:conformalembed}(B). The norm of the target Beltrami differential is shown in Figure \ref{fig:face2qcexample1}(A). The quasi-conformal parameterization is shown in (B). The histogram of the error between the target and output Beltrami differential is shown in (C), which indicates the obtained quasi-conformal parameterization is accurate. (D) shows the energy versus iterations.

\subsection{Computational time and comparisons}
Table \ref{table:computation} reports the computational details of our proposed algorithm. For meshes with about 10k faces, our proposed algorithm can generally obtain the parameterization onto the punctured disk with in 10 seconds on average. All the parameterization results have no overlapping faces, which means the obtained parameterizations are indeed bijective.

\begin{table}
\begin{center}
\begin{tabular}{c||c|c|c|c|c|c|c|c}
 Mesh & face & vertice  & Time spent & mean(mu) & std(mu) & \# of flips \\
\hline
Frog & 10603 & 5918  & 6.05 s & 0.0124 & 0.0148 & 0 \\
Boy & 15164 & 8699 & 5.90 s & 0.0196 & 0.0214 & 0 \\
Cat & 7857 & 4411  & 5.06 s & 0.0247 & 0.0297 & 0 \\
Bird & 10965 & 6210 & 7.57 s & 0.0258 & 0.0378 & 0 \\
Car & 10267 & 5805  & 8.57 s & 0.0264 & 0.0282 & 0\\
\end{tabular}
\caption{Computational details of the proposed algorithm}
\label{table:computation}
\end{center}
\end{table}

We also compare our proposed algorithm with the Ricci flow (RF) method \cite{JinRicci} and the inverse curvature Ricci flow (IDRF) \cite{InverseDistance} method. The results are reported in Table \ref{table:comparison}. As shown in the table, our method is stable under different regularities of the triangulations. On the contrary, both RF and IDRF fail on some meshes without remeshing. After improving the regularity of the triangulation through a remeshing process, RF and IDRF can be used to parameterize the mesh onto the punctured disk. To compare the quality of the conformal parameterization, we compute the mean and standard derivation of the angle distortion. From the table, it can be observed that our proposed algorithm outperforms RF and IDRF.

\begin{table}
\begin{center}
\begin{tabular}{ l || c | c | c }
Mesh & Ordinary RF \cite{JinRicci} & IDRF \cite{InverseDistance} & QCMC \\
\text{ } & (mean/sd) & (mean/sd) & (mean/sd) \\ \hline
Frog  & 0.1907/0.1548 & 0.0164/0.0196 & 0.0129/0.0164 \\ \hline
Boy  & 0.1647/0.1263 & 0.0212/0.0288 & 0.0203/0.0241 \\ \hline
Cat & 0.1852/0.1545 & 0.0252/0.0284 & 0.0233/0.0255 \\ \hline
Bird & fail & fail & 0.0245/0.0401 \\ \hline
Car  & 0.1881/0.1488 & 0.0277/0.0311 & 0.0253/0.0270 \\ \hline
2hole\_face  & 0.2937/0.2625& fail & 0.0114/0.0351 \\ \hline
2hole\_face (remeshed) & 0.2402/0.2412 & 0.0097/0.0175 & 0.0051/0.0058 \\ \hline
3hole\_face &0.3079/0.2591 & fail & 0.0195/0.0516 \\ \hline
3hole\_face (remeshed) & 0.2281/0.2205 & 0.0137/0.0220 & 0.0057/0.0070 \\
\end{tabular}
\caption{Comparison of different algorithms.}
\label{table:comparison}
\end{center}
\end{table}

\section{Applications}\label{application}
In this section, we show some applications of the conformal and quasi-conformal parameterizations.

\subsection{Surface remeshing}
\begin{figure}[t]
\centering
\includegraphics[width=3.5in]{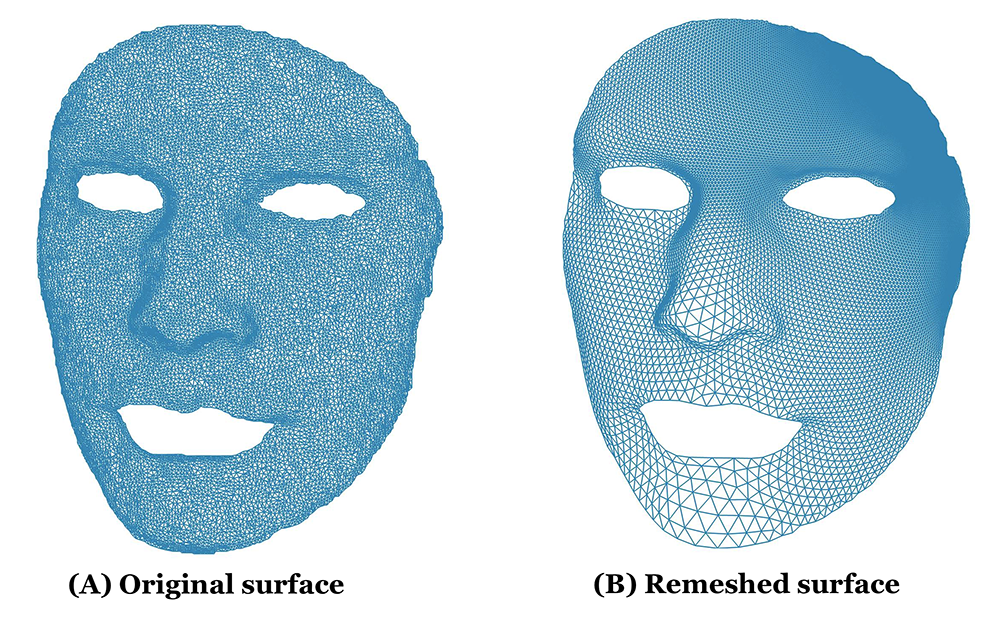}
\caption{(A) shows the original surface mesh. (B) shows the remeshed surface.\label{fig:remeshing1}}
\end{figure}

\begin{figure}[t]
\centering
\includegraphics[width=4.5in]{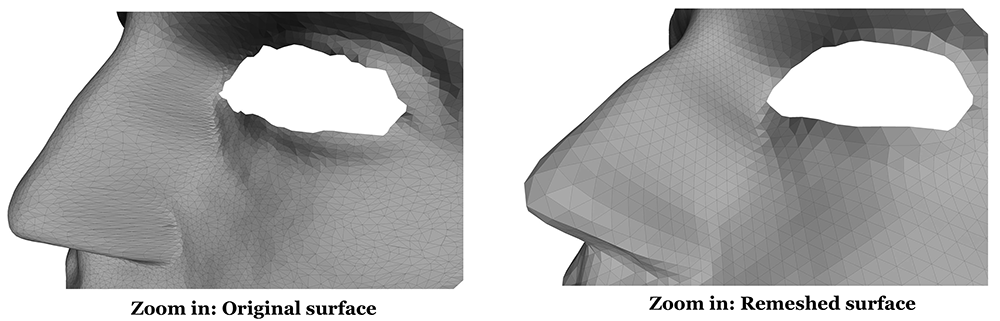}
\caption{The left shows the zoom in of the triangulation mesh of the original surface in Figure \ref{fig:remeshing1}(A). The right shows the zoom in of the triangulation mesh of the remeshed surface. \label{fig:remeshing2}}
\end{figure}

Surface remeshing refers to the process of improving the quality of the triangulation. This procedure is necessary in numerical computations to improve the accuracies of numerical solutions. A common technique to perform remeshing is done by parameterizing the surface onto a simple parameter domain (usually in $\mathbb{R}^2$). A regular mesh can be built on the simple parameter domain and the remeshing can be done through interpolation. Figure \ref{fig:remeshing1}(A) shows the original surface mesh. We conformally parameterize the surface onto a punctured disk and a regular mesh is built on the parameter domin. Remeshing is then done by interpolation. The remeshed surface is shown in (B). Figure \ref{fig:remeshing2}(left) shows the zoom in of the triangulation mesh of the original surface. The right shows the zoom in of the triangulation mesh of the remeshed surface. The quality of the triangulation is much improved.

\begin{figure}[!t]
\centering
\includegraphics[width=4.75in]{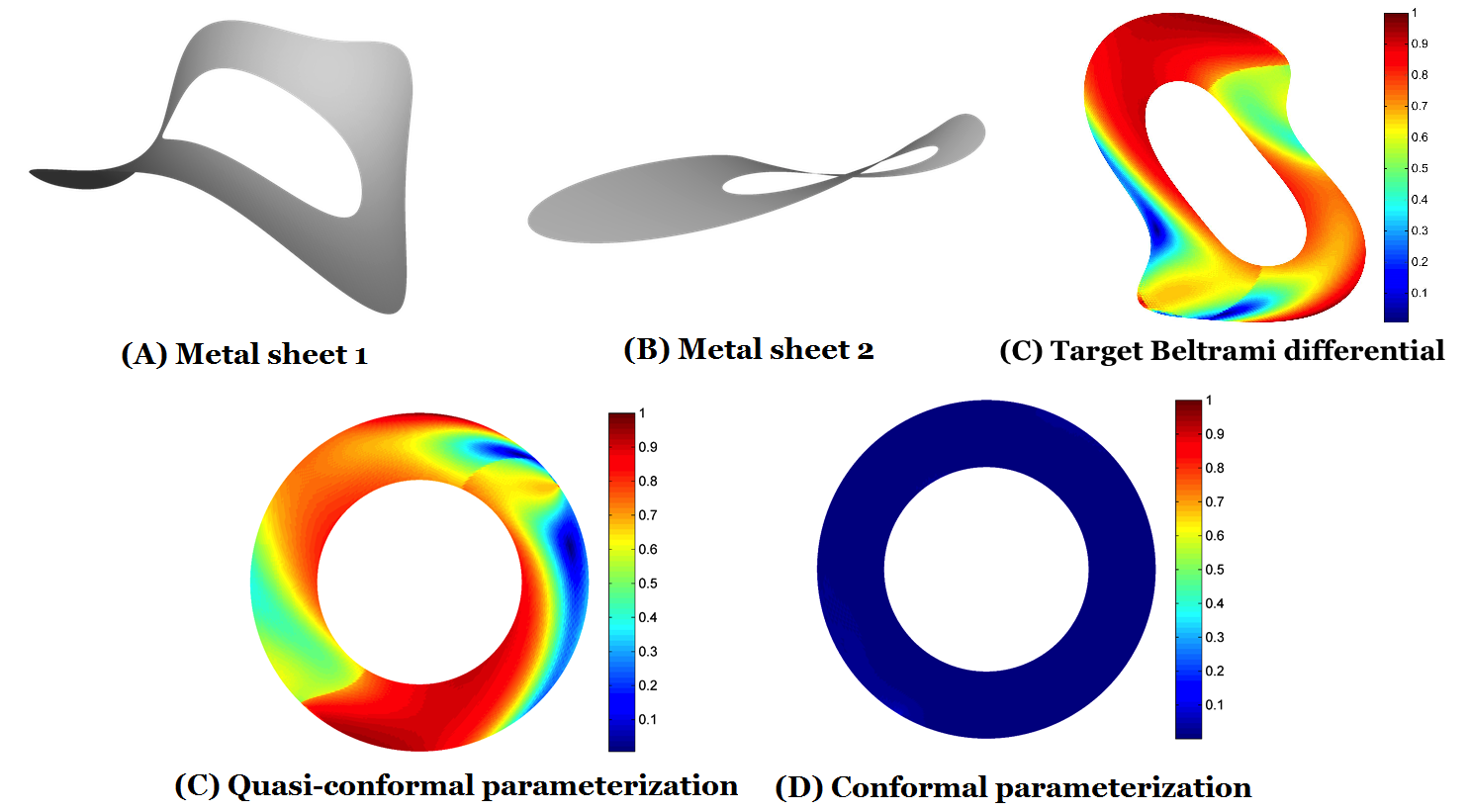}
\caption{(A) shows the metal sheet 1. (B) shows the metal sheet 2. (C) shows the quasi-conformal parameterization of the metal sheet 1. (D) shows the conformal parameterization of the metal sheet 2. The surface quasi-conformal map between the metal sheets can be obtained by the composition map.  \label{fig:metal}}
\end{figure}

\begin{figure}[t]
\centering
\includegraphics[width=3.75in]{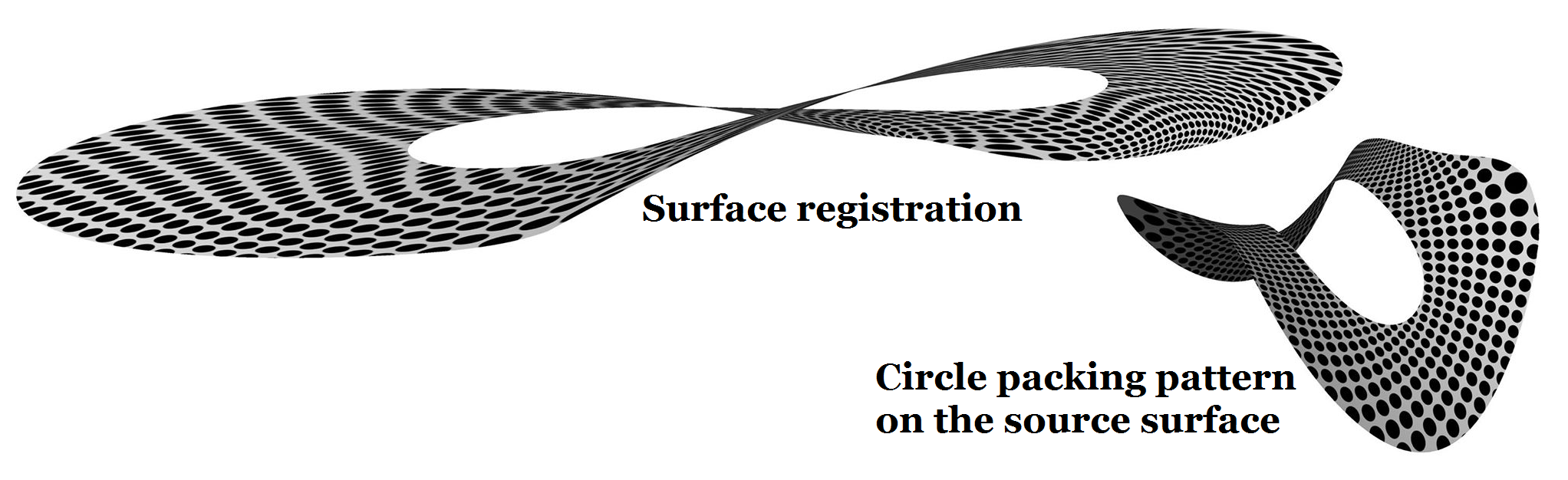}
\caption{Visualization of the surface quasi-conformal map between the metal sheets. The circle packing pattern on the original metal sheet is mapped onto the target metal sheet. Infinitesimal circles are mapped to infinitesimal ellipses.\label{fig:surfaceqc}}
\end{figure}

\subsection{Surface registration}
Quasi-conformal surface maps have been used for surface registration. Computing quasi-conformal surface maps between multiply-connected surfaces is challenging, due to the complicated topologies of the surfaces. By parameterizing the surfaces onto the punctured disk, the computation can be simplified. Figure \ref{fig:metal}(A) and (B) show two metal sheets, which are doubly-connected. Given a Beltrami differential on the metal sheet 1, our goal is to compute the quasi-conformal surface map between the metal sheets. To do this, we first quasi-conformally parameterize the metal sheet 1 with the prescribed Beltrami differential onto the annulus. The parameterization is as shown in (C) and the colormap is given by the norm of the Beltrami differential. Then, we parameterize the metal sheet 2 conformally onto an annulus, as shown in (D). In this way, the conformal parameter domains of metal sheet 1 and metal sheet 2 are conformally equivalent. Hence, their conformal modules are the same up to a M\"obius transformation. As shown in the figure, the inner radii of the two annulus are both approximately 0.6401. By the composition of the parameterizations, the quasi-conformal surface map between the two metal sheets can be obtained. Figure \ref{fig:surfaceqc} shows the visualization of the surface quasi-conformal map between the metal sheets. The circle packing pattern on the original metal sheet is mapped onto the target metal sheet. Infinitesimal circles are mapped to infinitesimal ellipses.

\begin{figure}[!t]
\centering
\includegraphics[width=4.65in]{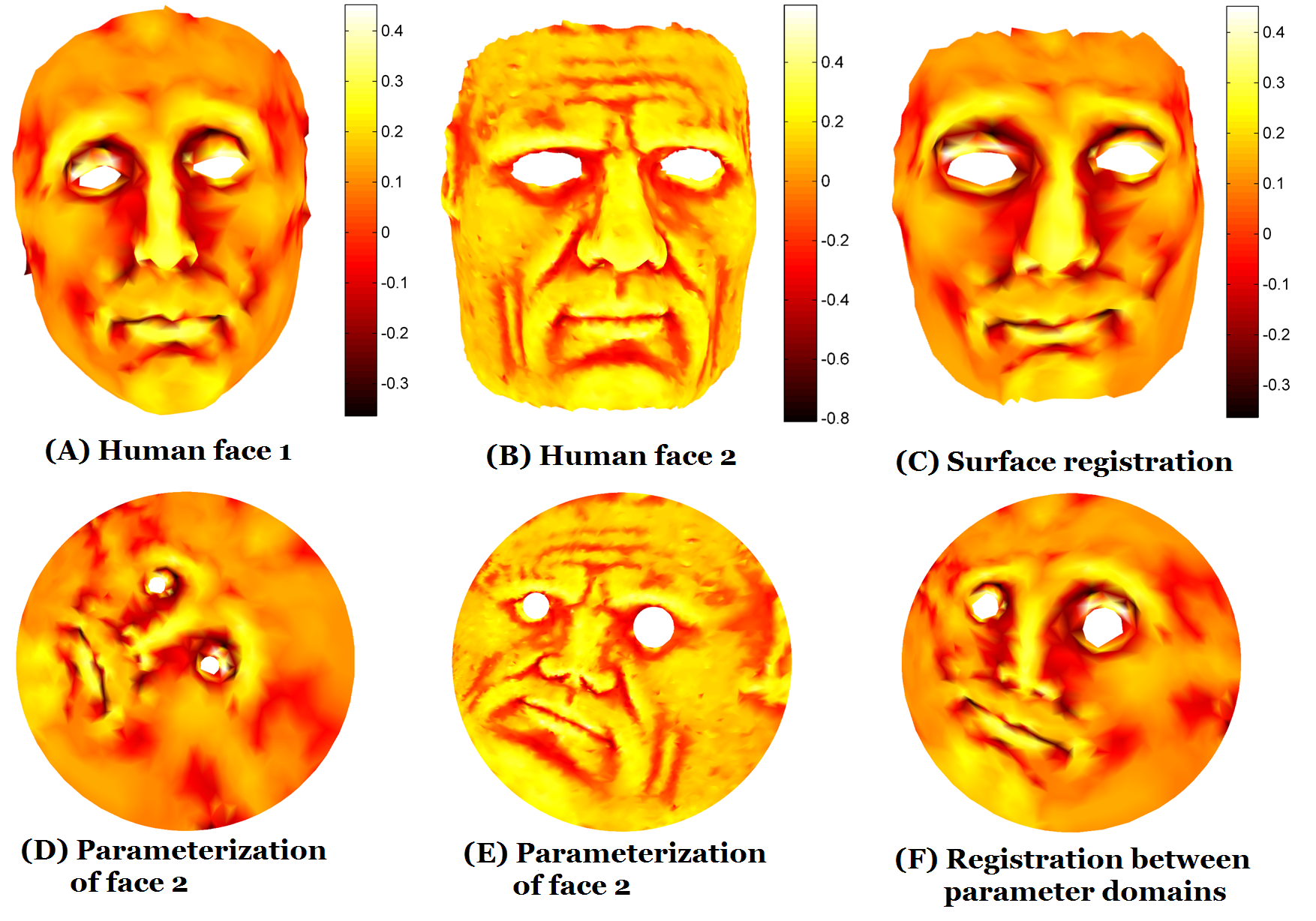}
\caption{(A) and (B) show the human face 1 and human face 2, whose colormaps are given by their mean curvature. (C) shows the surface registration between the human faces. (D) shows the conformal parameterization of face 1. (E) shows the conformal parameterization of face 2. (E) shows the registration between the conformal parameter domians.\label{fig:surfaceregistration}}
\end{figure}

Furthermore, by parameterizing the multiply-connected surfaces onto the punctured disk, surface registration can be computed easily on the simple parameter domains. Figure \ref{fig:surfaceregistration}(A) and (B) show two different human faces $S_1$ and $S_2$. We look for a geometric matching surface registration between them that matches curvatures. More specifically, we look for a diffeomorphism $f:S_1\to S_2$ that minimizes:
\begin{equation}
E_{geometric} (f) = \int_{S_1} |\mu(f)|^4 dS + \int_{S_1} (H_1 - H_2(f))^2 dS
\end{equation}
\noindent where $H_1$ and $H_2$ are the mean curvatures on $S_1$ and $S_2$ respectively. The first term minimizes the conformality (local geometric) distortion of the map $f$. The second term minimizes the curvature mismatching under the registration.

\begin{figure}[!t]
\centering
\includegraphics[width=4.65in]{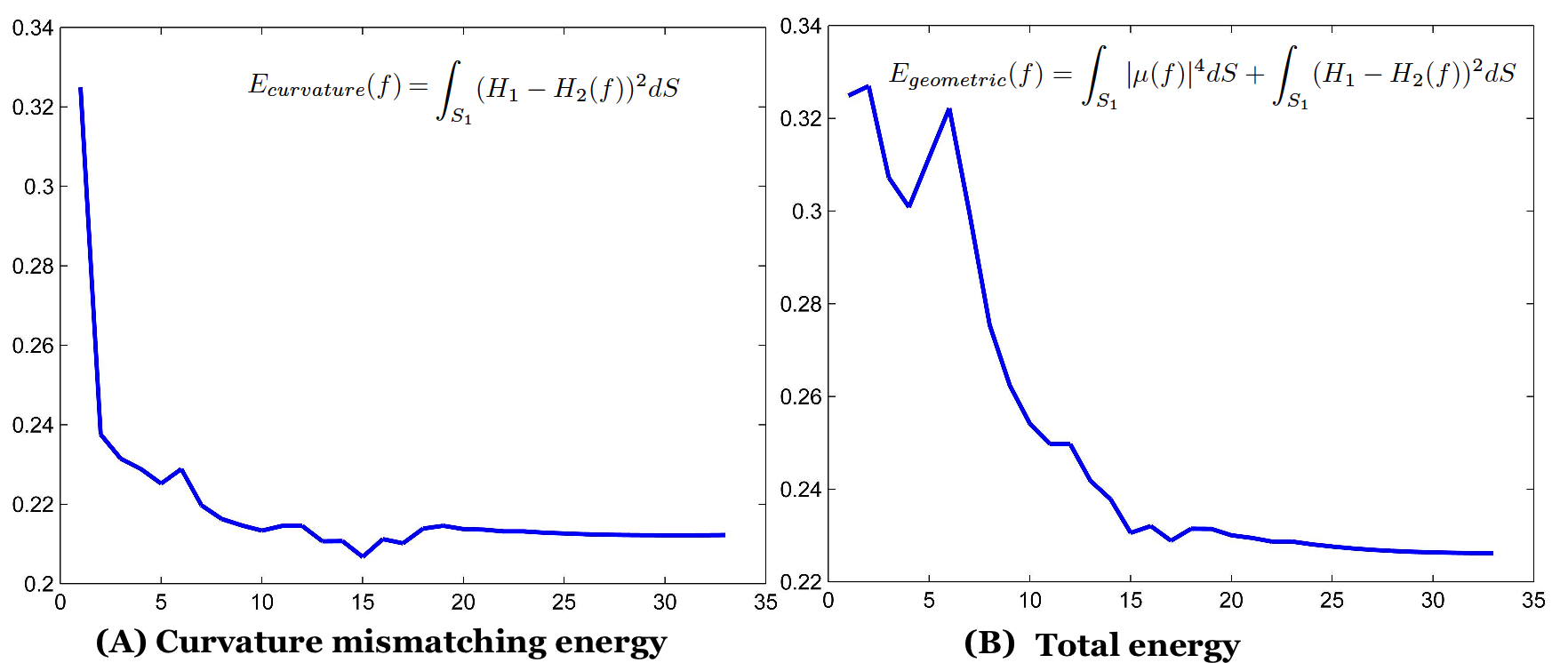}
\caption{Energy plots for the surface registration example. (A) shows the curvature mismatching versus iterations. It decreases as iteration increases until the optimal state is reached. (B) shows the total energy $E_{geometric}$ versus iterations. \label{fig:surfaceregistrationenergy}}
\end{figure}

We first conformally parameterize the two surfaces onto their canonical parameter domains, which are shown in (D) and (E). The colormaps on the two parameter domains are given by the mean curvatures of the two human faces. We then perform intensity-matching registration to find a mapping between the parameter domains that matches the curvature intensities. Using the obtained mapping, we transform source parameter domain to the target conformal parameter domain. The transformed domain is shown in (E). Finally, the surface registration can be obtained from the composition map. In (C), we map the curvature on the source surface to the target surface using the obtained surface registration. Note that the corresponding regions (high curvature regions) are consistently matched. It indicates that the obtained registration is geometric matching. Figure \ref{fig:surfaceregistrationenergy} shows the energy plots versus iterations during the process of registration between the parameter domains. (A) shows the curvature mismatching energy $E_{curvature}$ versus iterations. It decreases as iteration increases until the optimal state is reached. (B) shows the total energy $E_{geometric}$ versus iterations. 

\subsection{Shape signatures of multiply-connected objects}
\begin{figure}[t]
\centering
\includegraphics[width=4.65in]{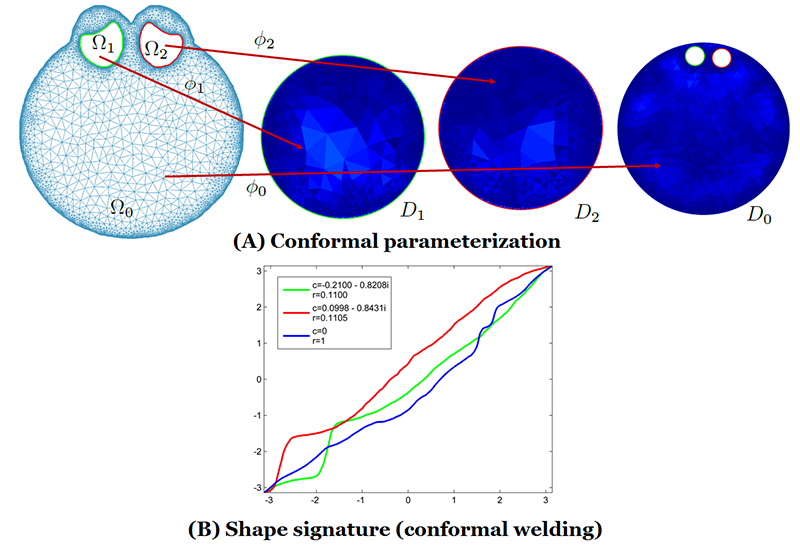}
\caption{(A) shows the conformal parameterizations of different sub-domains of the frog mesh. Using the conformal parameterizations, conformal weldings can be defined. The conformal weldings together with the conformal module define the shape signature of the frog mesh, which are shown in (B). \label{fig:shapesignature1}}
\end{figure}

Conformal parameterization of the 2D multiply-connected domain can be used to compute the shape signature representing the 2D shape. With the conformal parameterizations, conformal weldings can be computed, which can be used to define the shape signature. Figure \ref{fig:shapesignature1}(A) shows the frog mesh. It consists of three sub-domains, namely, $\Omega_0$, $\Omega_1$ and $\Omega_2$, together with the infinite domain $\Omega_{outside}$ outside the outermost boundary. Their conformal parameterizations, $\phi_0:\Omega_0\to D_0$, $\phi_1:\Omega_1\to D_1$, $\phi_2:\Omega_2\to D_2$ and $\phi_{outside} :\Omega_{outside}\to \mathbb{D}$, are computed. The conformal weldings can be obtained from the conformal parameterizations. More precisely, the conformal weldings can be computed as follows: $f_{01}:= \phi_1\circ\phi_0^{-1}:\mathbb{S}^1\to \mathbb{S}^1$, $f_{02}:= \phi_2\circ\phi_0^{-1}:\mathbb{S}^1\to \mathbb{S}^1$ and $f_{outside} := \phi_{outside}\circ \phi_0^{-1}:\mathbb{S}^1\to \mathbb{S}^1$. The conformal wedings together with the conformal module of $D_0$ form the shape signature of the frog mesh, as shown in (B).

\begin{figure}[!t]
\centering
\includegraphics[width=4.65in]{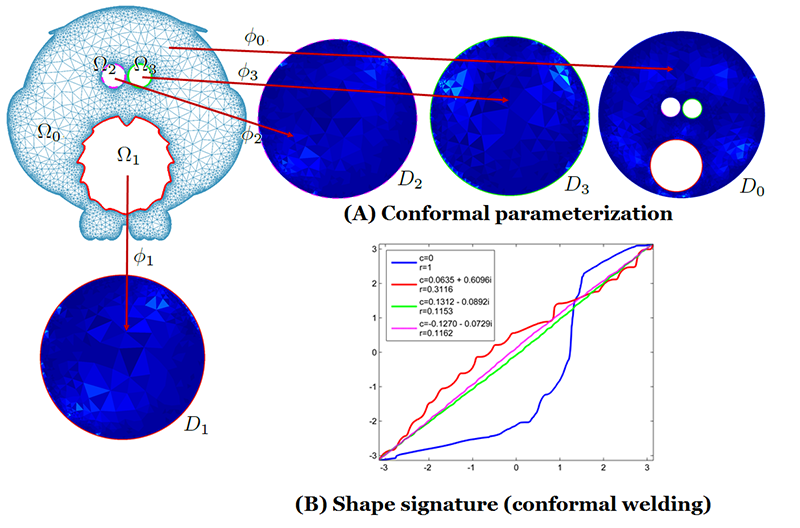}
\caption{(A) shows the conformal parameterizations of different sub-domains of the bird mesh. Using the conformal parameterizations, conformal weldings can be defined. The conformal weldings together with the conformal module define the shape signature of the bird mesh, which are shown in (B).\label{fig:shapesignature2}}
\end{figure}

Figure \ref{fig:shapesignature2}(A) shows the bird mesh and the conformal parameterizations of different sub-domains of the bird mesh. Using the conformal parameterizations, conformal weldings can be defined. The conformal weldings together with the conformal module of $D_0$ define the shape signature of the frog mesh, which are shown in (B). 

\section{Conclusion}\label{conclusion}
We address the problem of finding the quasi-conformal parameterization (QCMC) for a multiply-connected 2D domain or surface embedded in $\mathbb{R}^3$. Our goal is to map the multiply-connected domain one-to-one and onto a simple paramter domain. According to the quasi-conformal Teichm\"uller theory, given a prescribed Beltrami differential measuring the conformality distortion, a multiply-connected domain can be quasi-conformally parameterized onto a punctured disk. The center and radii of the inner circles, which is called the confomral module, can be uniquely determined up to a M\"obius transformation. In this paper, we propose an iterative algorithm to simultaeneously look for the conformal module and the optimal quasi-conformal map. The key idea is to minimize the Beltrami energy subject to boundary constraints. By incorporating the conformal module into the energy functional of the optimization problem, the quasi-conformal map and the conformal module can be simultaneously optimized. The parameterization of the multiply-connected domain simplifies numerical computations and has important applications in various fields, such as in computer graphics and visions. Experiments have been carried out on synthetic data together with real multiply-connected Riemann surfaces. Results show that our proposed method can efficiently compute quasi-conformal parameterizations of multiply-connected domains and outperforms other state-of-the-art algorithms. Applications of the parameterization technique have also been shown.

\end{document}